\documentclass[preprint]{emulateapj}
\usepackage{natbib}
\usepackage{graphicx}
\usepackage{epsfig}
\usepackage{subfigure}
\slugcomment{Submitted April 2, 2013; Accepted May 21, 2013}

\shorttitle{Extinction and PAH Intensity Variations in IRAS 12063-6259}
\shortauthors{D. J. Stock et al.}

\begin{document}

\title{Extinction and PAH Intensity Variations across the H~{\sc ii} region IRAS 12063-6259}

\author{D. J. Stock\altaffilmark{1}, E. Peeters\altaffilmark{1,2}, A. G. G. M. Tielens\altaffilmark{3}, J. N. Otaguro\altaffilmark{1} and A. Bik\altaffilmark{4} }

\altaffiltext{1}{Department of Physics and Astronomy, University of Western Ontario, London, ON, N6A 3K7, Canada}
\altaffiltext{2}{SETI Institute, 189 Bernardo Avenue, Suite 100, Mountain View, CA 94043, USA}
\altaffiltext{3}{Leiden Observatory, Leiden University, P.O. Box 9513, NL-2300 RA, The Netherlands}
\altaffiltext{4}{Max-Planck-Institut f\"{u}r Astronomie, K\"{o}nigstuhl 17, 69117 Heidelberg, Germany}

\begin{abstract}
The spatial variations in PAH band intensities are normally attributed to the physical conditions of the emitting PAHs, however in recent years it has been suggested that such variations are caused mainly by extinction. To resolve this question, we have obtained near-infrared (NIR), mid-infrared (MIR) and radio observations of the compact H {\sc ii} region IRAS 12063-6259. We use these data to construct multiple independent extinction maps and also to measure the main PAH features (6.2, 7.7, 8.6 and 11.2 \micron) in the MIR.  Three extinction maps are derived: the first using the NIR hydrogen lines and case B recombination theory; the second combining the NIR data with radio data; and the third making use of the Spitzer/IRS MIR observations to measure the 9.8 \micron\ silicate absorption feature using the Spoon method and PAHFIT (as the depth of this feature can be related to overall extinction). The silicate absorption over the bright, southern component of IRAS 12063-6259 is almost absent while the other methods find significant extinction. While such breakdowns of the relationship between the NIR extinction and the 9.8 \micron\ absorption have been observed in molecular clouds, they have never been observed for H {\sc ii} regions.  We then compare the PAH intensity variations in the Spitzer/IRS data after dereddening to those found in the original data. It was found that in most  cases, the PAH band intensity variations persist even after dereddening, implying that extinction is not the main cause of the PAH band intensity variations. 
\end{abstract}


\section{Introduction}

The mid-IR spectra of many Galactic and extragalactic sources display the well-known Unidentified Infrared (UIR) bands at 3.3, 6.2, 7.7, 8.6, 11.2, and 12.7 \micron\ (c.f. \citealt{1973ApJ...183...87G}; \citealt{1989ApJ...341..278G}; \citealt{1989ApJ...341..246C}). These broad emission features are generally attributed to the IR-relaxation of UV-pumped Polycyclic Aromatic Hydrocarbon (PAH) molecules (\citealt{1989ARA&A..27..161P}; \citealt{1989ApJS...71..733A}, \citealt{2008ARA&A..46..289T}). 

Observationally, the relative strengths, peak position, and profile of the PAH features have been seen to vary considerably within a source and from source to source (e.g. \citealt{2002AA...390.1089P}; \citealt{2001A&A...370.1030H}; \citealt{2004ApJ...611..928V}). The relative strengths of PAH features also change drastically going from neutral molecules to ions in laboratory studies (c.f. \citealt{1999ApJ...511L.115A}). In particular, the 3.3 and 11.2 \micron\ features arise primarily from neutral PAH molecules while the 6.2, 7.7, and 8.6 \micron\ features are attributed to PAH ions. This allows ratios involving the neutrals and ions, such as the 7.7/11.2 ratio, to be useful tools for probing the ionization state of the emitting molecules and hence for probing the local physical conditions (e.g. \citealt{2008ApJ...679..310G}). In addition, smaller molecules dominate the emission at shorter wavelengths while larger molecules emit predominantly at longer wavelengths \citep{1989ApJS...71..733A, 1993ApJ...415..397S}.

Despite the success of PAH ratios as tracers of physical conditions within a source, dust extinction often hampers measuring and interpreting these ratios. While there is extinction across the whole IR range, the main issue for PAHs is silicate extinction in the mid-IR which primarily affects the 8.6 and 11.2 \micron\ PAH bands. Moreover, this extinction can vary spatially across a source, complicating the interpretation of the observed mid-IR PAH features. \citet{2008ApJ...679..310G} investigated variations in the PAH ratios in a variety of objects including M82, and concluded that PAH variations are being caused by the properties of the emitting PAH population within the objects in their study. However, \citet{2008ApJ...676..304B} attribute the PAH band variations in this source to extinction variations. Indeed they posited that the variations in dust extinction are strong enough to affect the interpretation of their calculated PAH ratios, which showed variations comparable to the variations in extinction. 

This presents a fundamental problem in interpreting PAH ratios: are the variations in the band ratios intrinsic to the source? Are they the direct result of variations in the extinction due to dust? Or are they some combination of the two? Unfortunately, the only source in which this effect has been studied, M82, is a very complex and distant source, so each Spitzer/IRS pixel contains unresolved structure which adds to the already considerable uncertainties on the extinction and PAH measurements. On the other hand, many Galactic H{~\sc ii} regions can be spatially resolved at many wavelengths and the general principles of their structures are well understood (e.g. \citealt[][Chapter 7]{2005pcim.book.....T}). These properties greatly increase the precision of which the spatial variations of extinction can be studied within the target as well as giving access to multiple other methods of extinction measurement. As a result, studying the effects of extinction on PAH variations in a Galactic H{~\sc ii} region instead of M82 should result in a prime testbed to address this question.

Understanding the characteristics of dust in regions of massive star formation is of key importance as dust extinction and reemission dominates the spectral energy distribution of these regions. Studies on the properties of dust in H~{\sc ii} regions around massive young stars have a long history dating back to the optical studies on light scattering by \citet{1977ApJ...217..425M}, infrared studies by \citet{1972MNRAS.160....1W}, and sub millimeter \citet{1976ApJ...209...94W}. The advent of moderate resolution infrared spectrometers with wide coverage on ground-based and space based platforms has opened up novel ways of probing the dust characteristics in H~{\sc ii} regions through comparisons of the strength of IR recombination lines and studies of this type have revealed distinct differences with dust properties in the diffuse ISM (e.g. \citealt{1996A&A...315L.269L}). The absorption characteristics of dust in H~{\sc ii} regions can also be gleaned from detailed studies of the (infrared) spectral energy distribution (\citealt{2012ApJ...749L..21S}; \citealt{2011AJ....142....4S}; \citealt{2010A&A...518L..82K}). The high sensitivity of the IRS spectrometer on the Spitzer Space Telescope has opened up the study of the characteristics of dust in molecular clouds through photometric and spectroscopic studies of background stars shining through the cloud (\citealt{2007ApJ...666L..73C}; \citealt{2011ApJ...731....9C}; \citealt{2011A&A...526A.152V}; \citealt{2011ApJ...729...92B}). These observations reveal that the dust properties vary widely within molecular clouds and regions of star formation. These variations likely reflect the importance of coagulation of interstellar dust in larger and larger aggregates as the density increases (\citealt{2007A&A...461..215O}; \citealt{1994A&A...291..943O}).

There are several well known ways to measure the extinction in an H{~\sc ii} region of which three will be employed in this paper. The first two require observations of the ionized hydrogen emission from the center of the H{~\sc ii} region. We will employ the recombination lines in the NIR and this shall be referred to as the NIR method. The second method, the radio method, involves comparing the strength of the NIR hydrogen lines with the radio continuum generated by the same ionized hydrogen. The third method is independent of the first two, and fits the 9.8 \micron\ silicate profile to the observed spectra, deriving the depth of the 9.8 \micron\ silicate absorption feature; this will be referred to as the `Spoon' method \citep{2007ApJ...654L..49S}. In addition, the code PAHFIT does a similar analysis to the Spoon method in fitting the MIR spectrum and provides
 the optical depth of the  9.8 \micron\ silicate absorption feature.

We have therefore obtained narrow band photometric observations of the Paschen $\beta$ (hereafter Pa$\beta$) and Brackett $\gamma$ (hereafter Br$\gamma$) hydrogen lines of the Galactic H~{\sc ii} region (IRAS 12063-6259, He 2-77). We will combine these data with mid-IR Spitzer/IRS spectral maps and radio to create multiple extinction maps to ensure consistency. Subsequently, we can use these extinction maps to correct our Spitzer/IRS observations in order to investigate the effects of extinction on the PAH band ratios.

Section 2 will be devoted to background information about IRAS 12063-6259, including the relevant prior observations. In Sections 3 the data reduction processes relevant to the ISAAC, ATCA and Spitzer/IRS data are described. Section 4 shows the derivations of extinction maps from these data along with relevant discussion. In Section 5, the measurement of the PAH bands and their correlations in the reddened and dereddened data are presented. In Section 6, the morphology of IRAS 12063-6259 and the effects of extinction on PAHs is discussed and Section 7 presents conclusions. 

\section{IRAS 12063-6259}\label{sec:source}

\begin{figure}
	\begin{center}
	\includegraphics[width=8cm]{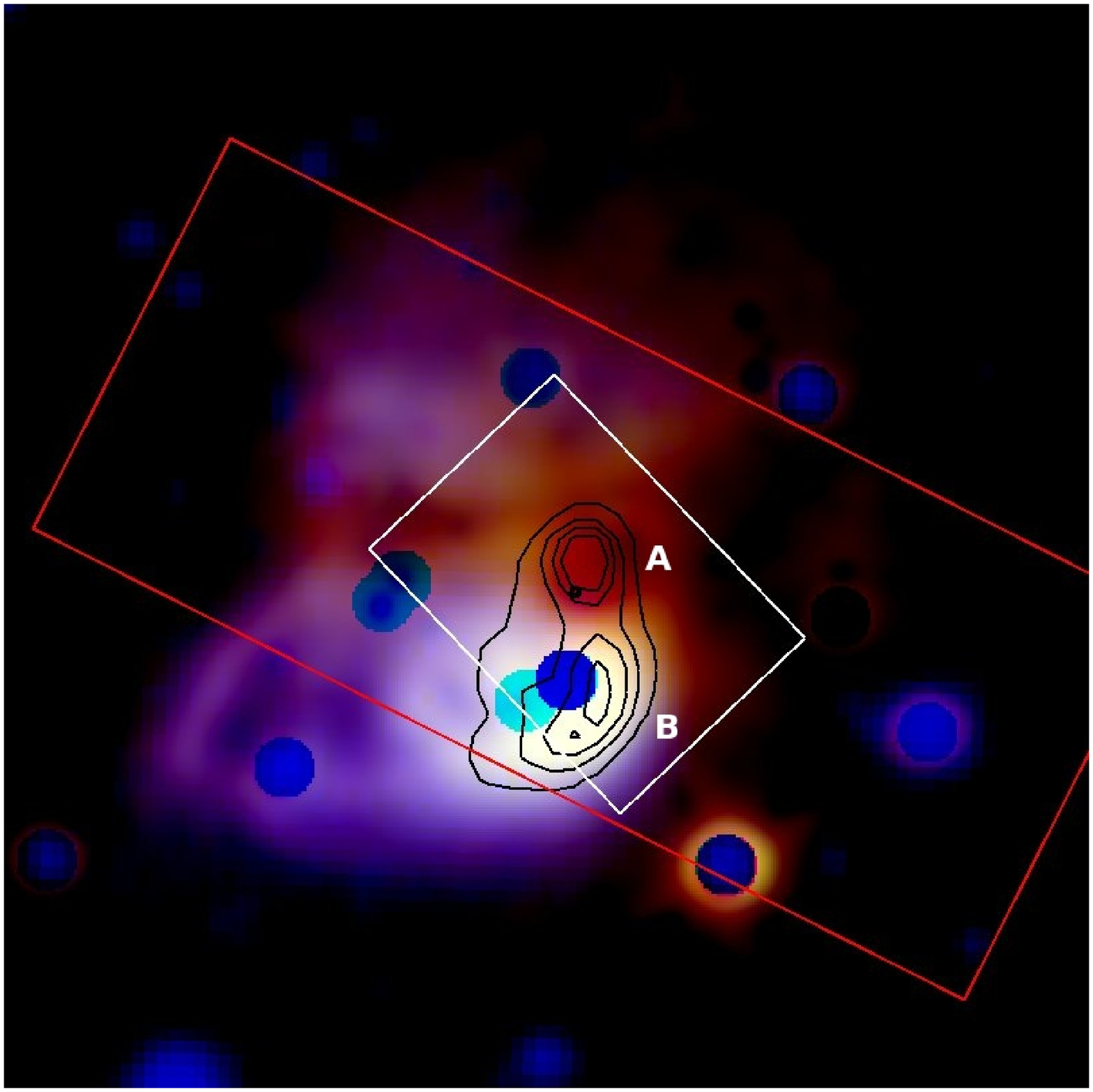}
	\end{center}
	\caption{Three color image of IRAS 12063-6259 with 8.6 GHz radio contours \citep{2003AA...407..957M} overlain in black with the two radio sources A \& B indicated. The image covers an area of 1' by 1', the red and green data are the background subtracted ISAAC 2.17 \micron\ Br$\gamma$ and 1.28 \micron\ Pa$\beta$ observations respectively and blue represents H$\alpha$ from the SHS (Southern hemisphere H$\alpha$ Survey, \citealt{2005MNRAS.362..689P}). The ISO-SWS aperture is shown in white and the IRS-SL cube FOV is shown in red. North is up and east is to the left. The dark circles are masked out areas around field stars. The dark lane across the center of the nebulosity is consistent with the observations of strong silicate 9.8 \micron\ absorption in that area (see Section~\ref{sec:sil_abs}).  }
	\label{fig1}
\end{figure}

IRAS 12063-6259, an ultra compact H~{\sc ii} region at a distance of 10 kpc \citep{1987AA...171..261C}, was first classified as a possible Planetary Nebula (PN) by \citet{1967ApJS...14..125H}. The first suggestions that it was perhaps not a PN were prompted by inspection of the spectral energy distribution (SED) in which the 10 \micron\ photometry was consistent with silicate absorption -- suggesting that it was an H~{\sc ii} region \citep{1980ApJ...238..585C}. This mis-classification is, in some ways, fortunate though, as its inclusion in PNe catalogs has provided abundances based on optical spectroscopy (e.g. \citealt{1994MNRAS.271..257K}). It was also observed in the radio in PNe surveys, revealing a very bright object at 5 GHz \citep{1979A&AS...36..227M}. 

Further radio observations by \citet{2003AA...407..957M} using the ATA array at 4.8 and 8.6 GHz indicate separation into three distinct sources, A and B with the B source further divided into B1 and B2 (A: 12:09:01.12 -63:15:52.9; B: 12:09:01.01 -63:16:00.8). It is thought that this separation reflects the likely stellar content - i.e. a cluster of three stars rather than a single ionizing star. We show an optical/NIR image of IRAS 12063-6259 in Figure~\ref{fig1} with radio contours. The dark lane in the center of Figure~\ref{fig1} indicates a high degree of extinction.

As shown in Figure~\ref{fig1}, the radio A source is coincident with a dark lane across the nebula while the radio B source occurs at the same location as the brightest patch of nebulosity in H$\alpha$, Pa$\beta$ and Br$\gamma$. The optical/NIR image shows this band most clearly, particularly the H$\alpha$ image. This feature is evident in the extinction maps, particularly in the NIR case B recombination map and partially in the mid-IR silicate feature maps. It is clear that the dark band is caused by extinction, rather than a lack of flux in the shorter wavelength filters, because the band obscures the continuum radio emission from the radio A H~{\sc ii} region.

The extinction band is also present in 2MASS JHKs images, which closely match the morphology of the H$\alpha$ image presented in Figure~\ref{fig1}. In the long wavelength (8.6 \micron) Spitzer/IRAC images the dark lane present in optical/near infrared data is absent and we see bright IR sources at the positions of both Radio A and B. However, the radio A source is substantially fainter at 3.6 \micron\ than the radio B source(s). This observation is in agreement with the optical/NIR conclusion that source A suffers from very high extinction.

Spectroscopic observations in the optical and NIR have yielded nebular conditions and abundances. Electron densities have been derived using both IR and optical methods. \citet{1994MNRAS.271..257K} found a density of 3000$\pm$1050 cm$^{-3}$ using the optical [S~{\sc ii}] doublet.  The IR [O~{\sc iii}] 88/52\micron\ lines loosely agree, yielding a value of 1335$^{+484}_{-284}$ cm$^{-3}$ \citep{2002AA...381..606M}. There are few temperature estimates of this object because heavy extinction in the optical obscures the standard temperature sensitive lines. \citet{1987AA...171..261C} found a value of 6400 K using observations of NIR hydrogen lines, however they noted that this value was uncorrected with respect to galactocentric radius and its effect on metallicity and therefore temperature. \citet{1994MNRAS.271..257K} performed the correction and found $T_e$ = 8800 K. This figure was found to agree with the $T_e$ = 9000 K found by \citet{1987ASSL..135..185J} based on the ratio of [S~{\sc iii}] 6312\AA\ and 18.7 \micron\ fluxes.

\section{Observations \& Data Reduction}

\begin{table*}[t!]
\small
\caption{\label{log_tot}Log of Observations } 
\begin{center}

\subfigure[VLT/ISAAC]{
\begin{tabular}{ccccccc}
Filter & Date & Integration Time & AB & Target & Seeing \\
       &      &  DIT $\times$  NDIT & Cycles & & (\arcsec) \\
\hline
1.21 \micron & 28/02/2010 & 3.6s $\times$ 39 & 5 & IRAS 12063-6259 & 0.71\\
1.21 \micron & 28/02/2010 & 3.6s $\times$ 8  & 5 & HD 115115 \\
1.28 \micron & 01/03/2010 & 3.6s $\times$ 12 & 5 & IRAS 12063-6259 & 1.53\\
1.28 \micron & 01/03/2010 & 3.6s $\times$ 8  & 5 & HD 115115 \\
2.17 \micron & 01/03/2010 & 3.6s $\times$ 39 & 5 & IRAS 12063-6259 & 0.94\\
2.17 \micron & 01/03/2010 & 3.6s $\times$ 8  & 5 & HD 115115 \\
2.19 \micron & 01/03/2010 & 3.6s $\times$ 12 & 5 & IRAS 12063-6259 & 0.83\\
2.19 \micron & 01/03/2010 & 3.6s $\times$ 8  & 5 & HD 115115 \\
\end{tabular}
\label{log_tot3}
}


\subfigure[Spitzer/IRS]{
\begin{tabular}{cc|cc|ccc}
Date & cycles $\times$  & \multicolumn{2}{c|}{Pointings}  & \multicolumn{3}{c}{Step Size}         \\
     &  ramp time (s)   &  $\parallel$ & $\bot$         & $\parallel$ & (\arcsec) & $\bot$            \\
\hline
19/03/2006 & 20 $\times$ 6.29 & 1                     & 7                   & 3.0         &       & 3.6               \\
\end{tabular}
\label{log_tot2}
}

\subfigure[ATCA$^b$]{
\begin{tabular}{cccccc}
Frequency & Configuration & Date & Integration Time  \\
  (GHz)   &                &     & (Minutes) \\
\hline
4.8  & 6D   & 09/04/2000 & 80 \\ 
8.64 & 6D   & 09/04/2000 & 80 \\
4.8  & 1.5D & 11/04/2000 & 80 \\ 
8.64 & 1.5D & 11/04/2000 & 80 \\

\end{tabular}  
\label{log_tot1}
}

\end{center}

$^a$ $\alpha, \delta$ (J2000); units of $\alpha$ are hours, minutes, and seconds, and
units of $\delta$ are degrees, arc minutes, and arc seconds.\\
$^b$ See \citet{2003AA...407..957M} for details.\\

\end{table*}

\subsection{ISAAC Data Reduction}

IRAS 12063-6259 was observed using the ISAAC instrument \citep{1998Msngr..94....7M} on UT3 (\textit{Melipal}) of the VLT (program ID:  084.C-0569). A complete log of our ISAAC observations is presented in Table~\ref{log_tot3}. Pa$\beta$ and Br$\gamma$ narrow band filters at 1.27 and 2.17 \micron\ were employed to observe the emission line intensities across the nebula. Observations were also performed using adjacent narrow band filters (1.21, 2.19 \micron) in order to measure the continuum component of the nebular emission. A standard star at similar airmass (HD 115115) was observed in the same narrow-band filters subsequent to the observation of the H{\sc ii} region. 

These ISAAC observations were performed in nodding/jittering configuration, allowing optimal removal of the sky background for both the nebular and standard star images by the ESO reduction pipeline. Each science frame was dark subtracted and flat-fielded using calibration data obtained that evening. The ISAAC data reduction pipeline was found to produce images with slightly offset WCS (World Coordinate System) information to the order of 5--10\arcsec\ in each reduction. The images were subsequently aligned by selecting stars with known WCS coordinates from the 2MASS point source catalog and adjusting the WCS information of each frame to match these coordinates.

The standard star, HD 115115, has spectral type F2V with 2MASS J and Ks band magnitudes of 8.717 and 8.473 respectively. The ratio of flux to be expected in a narrow band filter to that of a broad band filter was found by integrating the convolution of a normalized F2V spectral energy distribution (SED) with the relevant filter profile over the wavelength range of each filter and then dividing these to provide the expected flux ratio. For example:

\begin{equation}
\frac{ F^{*}_{1.21} }{F^{*}_{J}} = \frac{\int_{1.21-}^{1.21+} f(\lambda) S_{1.21}(\lambda) d\lambda  }{ \int_{J-}^{J+} f(\lambda) S_{J}(\lambda) d\lambda }
\end{equation}

where $ ^{F^{*}_{1.21}} / _{F^{*}_{J}}$ is the ratio of fluxes expected between the two filters, $1.21-$ and $1.21+$ are the upper and lower limits of the 1.21 \micron\ narrow band filter, $f(\lambda)$ is the flux of the star as a function of wavelength and $S(\lambda)$ is the filter profile as a function of wavelength. 

Using this relationship, the standard star flux can then be related to the observed counts to obtain the flux per count necessary for calibration of our science images. In each case, the standard star was observed at roughly the same airmass as the science target and immediately following the science observation. The airmass corrections were integrated into the process of determining the flux per count. 

The choice of standard star was constrained heavily by airmass and position on the sky constraints, as such we have checked our calibration by repeating this process for each star in both standard and object fields for which we have both photometry and 2MASS magnitudes. This assumed that most stars in both fields are likely to be late-type and therefore differences in their SEDs are likely to be minor and will not influence the calibration. In this process, we carefully excluded outliers, which were either a) embedded within the source, b) either optical doubles or visual binaries for which the ISAAC pipeline had incorrectly measured the counts, or c) early type stars. It was also necessary to remove stars with high 2MASS magnitudes ($\ge$ 11), as our observations are much more sensitive than those of 2MASS and this introduces non-linearities at very low count levels. Fits to this data provide the flux per count and allow us to calibrate our images. The fits for each filter are consistent with those derived using just the standard star. 

Table~\ref{log_tot3} also lists the seeing for each image as derived from the PSFs of the stars in the field by the ESO ISAAC pipeline. Unfortunately, one of the set of four images of IRAS 12063-6259 (1.28 \micron) has approximately double the seeing (1.5\arcsec) of the other images (0.7--0.9\arcsec). In order to compensate, we smoothed the three other images to match the seeing of the worst image (Br$\gamma$). The four images of the source were then calibrated, and images of Br$\gamma$ and Pa$\beta$ created by subtracting the adjacent background images (1.21 and 2.19 \micron\ respectively). 

Following the calibration process, the field stars in the images were masked to prevent them from influencing later science results. 

The S/N of the resulting Br$\gamma$ and Pa$\beta$ maps varies across IR 12063. Both achieve a S/N of greater than 10 per pixel in the area around radio B, but the area around radio A is more variable, with the S/N of Pa$\beta$ and Br$\gamma$ dropping to around 0.5 and 5 per pixel respectively. 

\subsection{Spitzer/IRS Data Reduction}\label{sec:IRS}

IRAS 12063-6259 was observed with the Short-Low (SL) module of the Infrared Spectrometer (IRS; \citealt{2004ApJS..154...18H}) on-board the Spitzer Space Telescope (PID: 20517, AOR: 14798848). We have summarized these observations in Table~\ref{log_tot2}. Both the first (SL1, 7.4 - 14.5 \micron, 1.8\arcsec/pixel) and the second (SL2, 5.2 - 7.7 \micron, 1.8\arcsec/pixel) orders were used to create a spectral map, with spectral resolutions $^\lambda/_{\delta\lambda}$ between 64 and 128. These observations cover an area of 57\arcsec $\times$ 26\arcsec\ (see the red box in Figure~\ref{fig1}). The point spread function (PSF) is about 2 pixels, roughly 3.6\arcsec.

The spectral mapping data obtained were processed through the pipeline reduction software at the Spitzer Science Center (Version 18.18). The BCD images were cleaned using CUBISM \citep{2007PASP..119.1133S}, wherein any rogue or otherwise `bad' pixels were masked, firstly using the built in `AutoGen Global Bad Pixels' with settings `Sigma-Trim' = 7, `MinBad-Frac' = 0.5 and `AutoGen Record Bad Pixels' with settings `Sigma-Trim' = 7 and `MinBad-Frac' = 0.75 and subsequently by manual inspection of each cube. Cleaned data cubes corresponding to each spectral order were made in this way. Finally, the SL2 spectra were scaled to match the SL1 spectra in order to correct for the mismatches between the two order segments. The bonus order SL3 was used to determine the scaling factors, since it overlaps slightly with both SL1 and SL2. These scaling factors were typically 5\%, well in agreement with the typical values of 10\% obtained by \citet{2007ApJ...656..770S}. Spectra were extracted from the spectral maps by moving, in one pixel steps, a spectral aperture of 2x2 pixels in both directions of the maps. This results in overlapping extraction apertures.

Before measuring PAH fluxes in the IRS spectral cube, the continuum was removed from each spectrum by computing a spline polynomial fit to the continuum (including a point at 8.2 \micron) and then subtracting the spline (see e.g. \citealt{2001A&A...370.1030H}, \citealt{2002AA...390.1089P}). The most prominent features were measured by integrating all of the remaining flux between fixed points, e.g. 5.86 -- 6.6 \micron\ for the 6.2 \micron\ feature. The 7.7\micron\ complex was also measured using this method, while the weaker features (6.0, 11.0\micron) were fitted with gaussians. For 6.0 and 11.0 these were then subtracted from the 6.2 and 11.2 \micron\ fluxes to compensate for their inclusion in the initial integration. The 12.7 \micron\ feature appears blended with the 12.8 \micron\ [Ne~{\sc iii}] line along with the 12.3 \micron\ H$_2$ line, all of which were fitted concurrently in order to minimize uncertainties. For the gaussian fits, uncertainties are generated in the fitting process, while for the features for which the integration was used, uncertainties were calculated by measuring the rms noise in various wavelength windows and then combining this with each flux measurement to find the signal to noise ratio. 

\subsection{ATCA Data Reduction}\label{sec:atca}

\citet{2003AA...407..957M} observed IRAS 12063-6259 using the Australia Telescope Compact Array (ATCA, Program C868) in April/May 2000 using the 6 km (6D) and 1.5km (1.5D) configurations at 4.8 and 8.6 GHz. Details of the reduction of this dataset are discussed by \citet{2003AA...407..957M}. A log of the relevant observations is shown in Table~\ref{log_tot1}. The configuration of the ATCA array used for these observations means that sensitivity drops off away from the source and very little extended structure is detected. 

\section{Results and Analysis}

\subsection{NIR Hydrogen Lines}\label{sec:NIRH}

The general formula for differential extinction $A_{\lambda_1} - A_{\lambda_2}$ between two different wavelengths $\lambda_1$ and $\lambda_2$ is given by:

\begin{equation}
A_{\lambda_1} - A_{\lambda_2} = -2.5 \times \left[ log\left(\frac{F(\lambda_1)}{F(\lambda_2)}\right) - log\left(\frac{F_0(\lambda_1)}{F_0(\lambda_2) } \right)  \right]
\label{eq:1}
\end{equation} 

where $F_0(\lambda_1)$ is the intrinsic flux and $F(\lambda_1)$ is the observed flux of a line at $\lambda_1$. Given an electron density of 1000--3000 cm$^{-3}$ and temperature of $\sim$ 8800 K (see Section~\ref{sec:source}), a linear interpolation between the nearest case B grid points provided by \citet{1987MNRAS.224..801H} gives $\frac{F_0(Pa\beta)}{F_0(Br\gamma)} = 5.82$. Equation~\ref{eq:1} can then be restated for this specific case as:

\begin{equation}
A_{Pa\beta} - A_{Br\gamma} = -2.5 \times log\left(\frac{F(Pa\beta)}{5.82 \times F(Br\gamma)}\right) 
\end{equation} 

Extinction in the NIR ($\sim$ 1--3 \micron) is described by a power law, usually taken to be of the following form \citep{1990ARAA..28...37M, 1990ApJ...357..113M}:

\begin{equation}
A_\lambda = \frac{A_K}{(\lambda/2.2)^{\alpha}}
\label{eq:AK}
\end{equation}

where $\lambda$ is in \micron\ and the exponent $\alpha$ is usually in the range of 1.7--2.0 for the ISM (e.g. \citealt{1990ApJ...357..113M}).

It is then possible to state both $A_{Pa\beta}$ and $A_{Br\gamma}$ in terms of $A_K$ by substituting Eq~\ref{eq:AK} as appropriate:

\begin{equation}
A_K = - \frac{2.5}{1.62} \times log\left(\frac{F(Pa\beta)}{5.82 \times F(Br\gamma)}\right) 
\end{equation}

adopting $\alpha$ = 1.8 (e.g. \citealt{2002AA...381..606M}).

\begin{figure}
	\begin{center}
	\includegraphics[width=8cm]{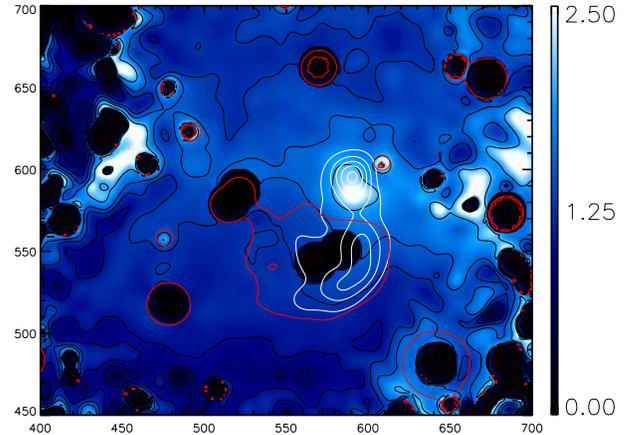}
	\end{center}
	\caption{Extinction map derived using case B recombination assumptions and ISAAC narrow band hydrogen line images. Black contours represent A$_K$ of 0.25, 0.5, 0.75, 1.0, 1.25, 1.5, 2; white contours represent the 4.8 GHz radio observations (as in Figure~\ref{fig1}). The red contours surrounding radio B and some bright stars represent the Pa$\beta$ S/N level of 3. Solid black circles represent areas in which foreground stars have been masked out. The $x$ and $y$ axes refer to positions in pixels on the NIR images (0.148\arcsec\ per pixel).}
	\label{fig2}
\end{figure}

Upon substitution of our flux calibrated maps, this produces the extinction map shown in Figure~\ref{fig2}. The area of highest extinction is correlated with the dark lane across the nebula (Figure~\ref{fig1}). The highest extinction values ($A_K \sim 2$) are observed at the same position as the radio A source (and as such are compromised by a lack of signal to noise), the second radio source (radio B) is obscured to a much lower extent ($A_K \sim 1 - 1.5$). 

The derived values of $A_K$ (0.6 -- 2.5) are consistent with the average value of 0.8 derived by \citet{2002AA...381..606M} using ISO-SWS data. The \citet{2002AA...381..606M} value would be biased towards the lower values as the ISO aperture (14\arcsec x 20\arcsec, see Figure 2 of \citealt{2003AA...407..957M}) included the brightest (lowest extinction) parts which dominate the emission line spectrum and lead to a lower overall estimate of extinction. 

This method will henceforth be referred to as the NIR method.

\subsection{Radio Continuum and Near-IR Hydrogen Lines}

From radio observations of the ionized hydrogen continuum at multiple frequencies it is possible to derive the intrinsic intensity of recombination lines that are also emitted. This can be used to measure extinction as the observed flux of a hydrogen line (such as those discussed in the previous section) can be compared to the flux predicted by the radio observations. In order to derive the flux implied by the radio observations, some other quantities (such as the electron temperature and emission measure) must also be derived. Using the ATCA radio observations (discussed in Section~\ref{sec:atca}) a brightness temperature map for both 4.8 and 8.6 GHz observations can be calculated. The brightness temperature is defined as \citep[][equation 7.70]{2005pcim.book.....T}:

\begin{equation}
S_{\nu} = \frac{2 \nu^2}{c^2} k T_B \Omega 
\end{equation}

where $\nu$ is the frequency in Hz, $T_B$ is the brightness temperature in K and $\Omega$ is the source solid angle. The source solid angle has been explicitly included to facilitate the creation of brightness temperature maps (i.e. per pixel), rather than the usual method of using the sum of all radio flux for the whole object. 

We can rearrange to find:

\begin{equation}
T_B = 3.26 \times 10^{-5} \times \frac{S_{\nu}}{\nu_{GHz}^2 \Omega } \quad [K]
\label{eq:tb_deriv}
\end{equation}

We will refer to the maps created using this equation as the ``observed" brightness temperature maps. The brightness temperatures are in general much higher for the 4.8 GHz map, peaking at around 300 K as opposed to 50 K for the 8.64 GHz map. 

The unusual configuration of the ATCA array means that the ATCA beam is an elliptical rather than a circular aperture. As it is necessary to combine the two maps, they were combined by smoothing the higher frequency (8.64 GHz) map using the larger beam of the low frequency (4.8 GHz) map, taking into account this ellipticity.

The following equations then describe the relationship between the important quantities necessary for calculating the intrinsic emission of any hydrogen recombination line, namely the electron temperature, $T_e$ and emission measure, $EM$:

\begin{equation}
T_B = T_e ( 1 - e^{-\tau(\nu)} ) \quad [\textrm{K}]
\label{eq:tb} 
\end{equation}

and:

\begin{equation}
\tau(\nu) = (8.235 \times 10^{-2}) a T_e^{-1.35} \left(\frac{\nu}{GHz}\right)^{-2.1} \left(\frac{EM}{cm^{-6} pc}\right)
\label{eq:nu}
\end{equation}

where $\nu$ is the frequency of our radio observations, $\tau(\nu)$ is the optical depth for that frequency and $a$ is a function of $T_e$ and $\nu$ and represents the ratio of the true optical depth to a simplified approximation \citep{1967ApJ...147..471M}.

$T_e$ was then found using iterative calculations (cf. \citealt{1997ApJ...487..818W}; \citealt{2006agna.book.....O}) after combining Equations~\ref{eq:tb} and \ref{eq:nu} to find:

\begin{equation}
T_{B,4.8} = T_e \left( 1 - e^{-\tau_{8.64} \times \left[\frac{4.8}{8.64}\right]^{-2.1} }  \right) \quad [\textrm{K}]
\label{eq:fin}
\end{equation}

where $T_{B,4.8}$ is the observed 4.8 GHz brightness temperature map and the factor $\tau_{8.64} \times \left[\frac{4.8}{8.64}\right]^{-2.1}$ represents the optical depth at 4.8 GHz as implied by the 8.64 GHz measurements. For the first iteration, the temperatures measured by previous authors (as mentioned in Section~\ref{sec:source}) of around 9000 K was used for the whole map. The difference between the observed brightness temperature map and that calculated using Equation~\ref{eq:fin} was applied iteratively to correct our $T_e$ map until no further changes were observed. The resultant map shows that radio A has a slightly lower average temperature ($\sim$~8000 K) than radio B (9000 K) and that the temperature throughout most of the emission is roughly constant at around 9000 K. Knowing $T_e$, Equation~\ref{eq:nu} can then be rearranged to calculate the emission measure ($EM$) for each pixel:

\begin{equation}
EM =  4.72  a^{-1} T_e^{1.35} \left(\frac{\nu}{GHz}\right)^{2.1}  \tau(\nu)\quad [\textrm{cm}^{-5}]\textrm{.}
\label{eq:em}
\end{equation}

Following \citet{1997ApJ...487..818W}, we then calculate the expected flux at Br$\gamma$ using the following equation:

\begin{equation}
S_{Br\gamma} = 0.9 h\nu_{Br\gamma} \alpha^{eff}_{Br\gamma} \frac{\Omega}{4\pi} EM \quad [\mathrm{erg\ cm^{-2} s^{-1}}]
\label{eq:brg0}
\end{equation}

where it is assumed that the hydrogen abundance by number is 0.9, h is Planck's constant, $\nu_{Br\gamma}$ is the frequency of a $Br\gamma$ photon, $\Omega$ is the solid angle subtended by the emitting source and

\begin{equation}
\alpha^{eff}_{Br\gamma} = (6.48 \times 10^{-11}) T_e^{-1.06}
\end{equation}

represents the effective recombination cross section \citep{1987MNRAS.224..801H}.

To find a specific expression for $S_{Br\gamma}$, we can substitute Eq.~\ref{eq:em} into Eq.~\ref{eq:brg0} to yield:

\begin{equation}
S_{Br\gamma} = 4.24 h\nu_{Br\gamma} \alpha^{eff}_{Br\gamma} \nu_\mathrm{4.8 GHz}^{2.1} \frac{\Omega}{4\pi} T_e^{0.29} \tau(\nu) \quad [\mathrm{erg\ cm^{-2} s^{-1}}]
\label{eq:brg}
\end{equation}

which was calculated for each pixel of the map. The extinction A$_{Br\gamma}$, is then calculated by comparing the observed and calculated Br$\gamma$ fluxes (Equation~\ref{eq:AKrad}). 

\begin{equation}
A_{Br\gamma} = -2.5 \times log \left(\frac{S_{Br\gamma}}{S_{Br\gamma,0}}\right) 
\label{eq:AKrad}
\end{equation}

\begin{figure*}
	\begin{center}
	\includegraphics[width=18.5cm]{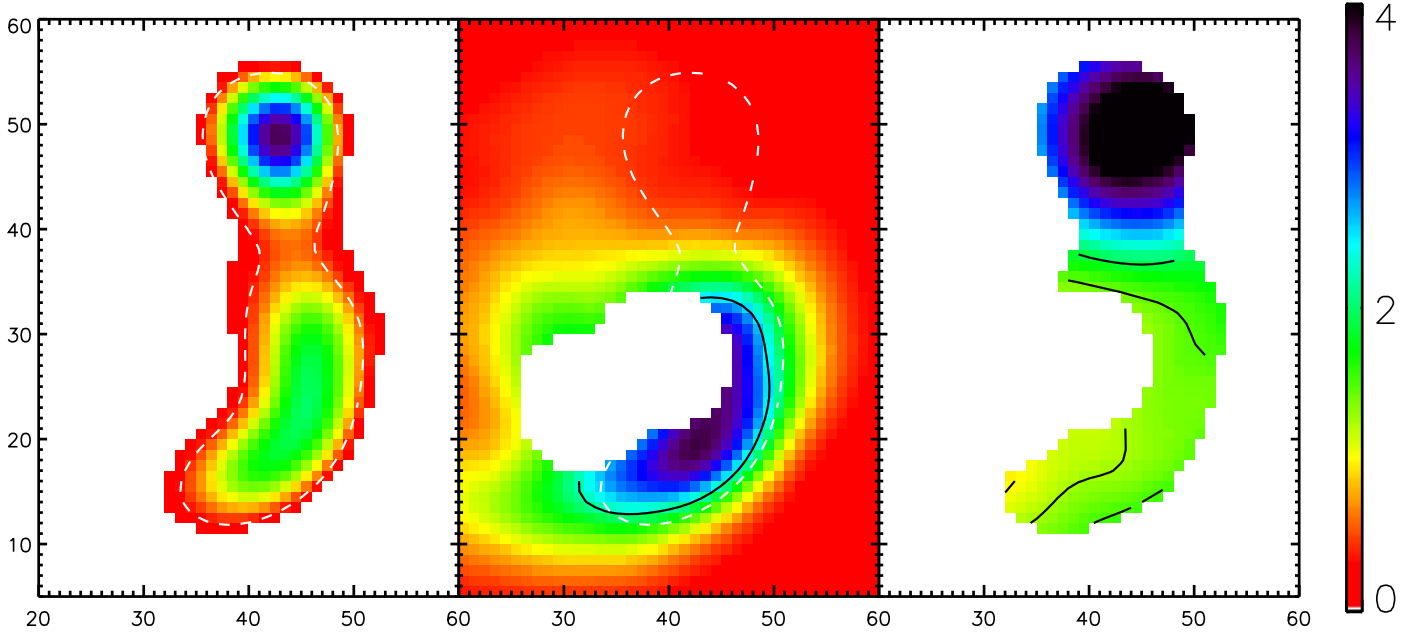}
	\end{center}
	\caption{Predicted Map of the Br$\gamma$ flux derived from ATCA radio observations (left, white dashed contour at 1.5 x 10$^{-13}$ ergs s$^{-1}$ cm$^{-2}$); Br$\gamma$ flux observed with ISAAC smoothed to approximately the same resolution as the radio data, where the white region is a masked out star (middle, black contour represents 5 x 10$^{-14}$ ergs s$^{-1}$ cm$^{-2}$), white contours are those of the data in the left panel) and the $A_K$ map resulting from Equation~\ref{eq:AKrad} (right, contours at $A_K$=0.5, 1, 2, 3). The colorbar to the right refers exclusively to the image in the right panel. For each panel, the $x$ and $y$ axes are positions in pixels on the radio image (0.3\arcsec\ per pixel).   }
	\label{fig5}
\end{figure*}

From this, the K band extinction, A$_K$, is found as before, using Eq.~\ref{eq:AK}. The calculated Br$\gamma$ map (Equation~\ref{eq:brg}), is shown along with the observed ISAAC Br$\gamma$ map smoothed to the same resolution, in Figure~\ref{fig5}, along with the resultant extinction map. This method will be referred to as the radio method.

\subsection{Mid-IR Silicate Absorption}\label{sec:sil_abs}

\begin{figure}
	\begin{center}
	\includegraphics[width=8cm]{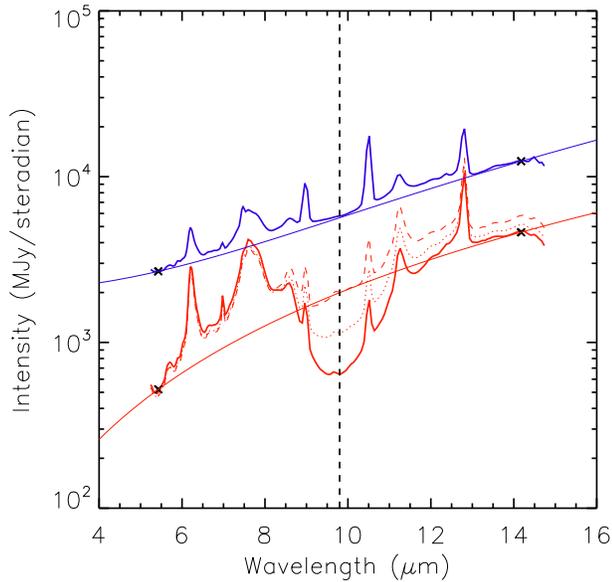}
	\end{center}
	\caption{Example of the Spoon method \citep{2007ApJ...654L..49S} using Spitzer/IRS spectra. The red and blue lines represent the Spitzer/IRS SL spectra of Radio A (Spoon A$_K$ = 2.3) and B (Spoon A$_K$ = 0.0) respectively. The radio B spectrum has been offset from the radio A spectrum for clarity. The thin red and blue lines are the initial interpolated power law continua for both cases (with the interpolation points marked with crosses), the black dashed line at $\lambda$ = 9.8 \micron\ shows the approximate location of the silicate feature. The red dotted and dashed lines are the spectrum of radio A after being dereddened using the non-iterative Spoon method (dotted) and the iterative Spoon method (dashed) and then normalized to the observed spectrum. }
	\label{fig6}
\end{figure}

\citet{2007ApJ...654L..49S} proposed a method for measuring the depth of the 9.8 \micron\ silicate absorption feature independent of any intrinsic profile of the absorption. The Spoon method, for PAH dominated spectra (see \citealt{2007ApJ...654L..49S}), is as follows: interpolate a power law continuum using points at 5.5 \micron\ and 14.5 \micron\ of the form $y = ax^k$, then calculate the natural log of the ratio of the flux of this continuum at 9.8 \micron\ and the observed flux at 9.8 \micron. This value, called $S_{sil}$ by \citet{2007ApJ...654L..49S}, can be interpreted as the optical depth at 9.8 \micron, $\tau_{9.8 \micron}$. An example of this method employed on the Spitzer/IRS spectra of Radio A and B is presented in Figure~\ref{fig6}. 

As discussed by \citet{2006ApJ...653.1129B}, if significant silicate extinction is present, the flux of the long wavelength continuum point will be affected by absorption because of the overlapping of the 9.8 and 18 \micron\ bands. \citet{2006ApJ...653.1129B} suggest that if a Long-Low (LL) IRS cube is available, that the 14.5 \micron\ continuum point can be abandoned in favor of one at 23 \micron\ which is less susceptible to silicate absorption. LL data for IRAS 12063-6259 are not available, so a new approach, iteratively calculating the extinction using the Spoon method in combination with a wavelength profile of the absorption, has been devised.

When the 14.5 \micron\ continuum point is affected by silicate absorption, the Spoon method underestimates the continuum at 9.8 \micron\ -- leading to an underprediction of $\tau_{9.8}$. Dereddening such IRS spectra using this value of $\tau_{9.8}$, will still leave unaccounted for silicate absorption (this is shown in Figure~\ref{fig6}, where the dashed red line is the result of dereddening the radio A spectrum in this way). The new method implemented here applies the Spoon method again to such a dereddened spectrum, to find a new value of $\tau_{9.8}$. This process of applying the Spoon method and dereddening the spectrum continues until the flux of the continuum point at 14.5 \micron\ ceases to change. The continuum point at 9.8 \micron\ in the final continuum is then compared with the observed 9.8 \micron\ flux in the original spectrum. For the spectra most affected by silicate absorption ($\tau_{9.8}$ $>$ 2.0), the difference between the initial value provided by the Spoon method and the final value was around 40\%. 

\begin{figure}
	\begin{center}
	\includegraphics[width=8cm]{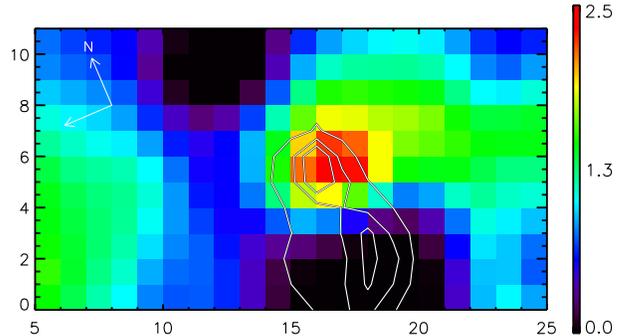}
	\end{center}
	\caption{Map of the 9.8 \micron\ silicate absorption as measured by the Spoon method. White contours represent 4.8 GHz brightness temperature 50, 100, 150 K. The $x$ and $y$ axes refer to positions in pixels on the IRS cube (1.8\arcsec\ per pixel). }
	\label{fig_sil}
\end{figure}

In practice, the \citet{2006ApJ...637..774C} description of the shape of the 9.8 and 18 \micron\ silicate extinction features in the ISM to is used to deredden the IRS SL spectra at each stage. The \citet{2006ApJ...637..774C} model also gives a direct relationship between A$_{9.8}$ and A$_K$, which turns out to be nearly equal to one (1.006). This provides the following relation between $\tau_{9.8}$ and A$_K$:

\begin{equation}
A_K = \frac{A_{9.8}}{1.006} =  \frac{1.086 \times \tau_{9.8}}{1.006} = 1.079 \times \tau_{9.8} \mathrm{.}
\label{eq:spoon}
\end{equation}
 
So to within 10\%, A$_K$ is equal to $\tau_{9.8}$ under the conditions for which the \citet{2006ApJ...637..774C} description of the silicate features in the local ISM are satisfied. This method will be referred to as the Spoon method in future discussion. The map of extinction as measured by the Spoon method is presented in Figure~\ref{fig_sil}. 

\subsection{Comparisons between Extinction Maps}\label{sec:extdisc}
Before using the various extinction maps to deredden mid-IR observations, they can be compared for consistency and correlations. In each case, the higher spatial resolution observation must be binned to match the resolution of the lower and then the derived extinctions for each pixel can be compared.
The three maps have pixels of size 0.15\arcsec, 0.3\arcsec\ and 1.8\arcsec\ for the NIR, radio and IRS data respectively. As such, S/N improves significantly upon binning. However, it is still likely that many NIR points will be untrustworthy as a) the native resolution measurements of Pa$\beta$ which are used to create them are have very low S/N and b) the morphology of the NIR narrow band emission does not match that of the MIR or radio emission. Point b also applies to the radio method, as the Br$\gamma$ image is used for this method. 
Morphologically, the only large scale comparison can be drawn between the NIR map and the Spoon map as they cover large areas. The NIR map (Figure~\ref{fig2}) shows the prominent bar ($A_K$ = 1.5 -- 2) running from east to west, with the highest extinction ($A_K$ = 2.5) occurring very near to the location of radio A within the bar. The southern regions, near radio B but also covering the extended emission evident in the H$\alpha$ map shown in Figure~\ref{fig1}, display an $A_K$ of between 1 and 1.25. In contrast, the Spoon map (Figure~\ref{fig_sil}) shows zero extinction in the southern regions near radio B. It agrees with the NIR map in that the peak extinction occurs near radio A (although it underestimates the degree of extinction), but the `bar' of extinction visible in the NIR is only somewhat present as it falls between the radio A peak and the large region of silicate absorption occurring to the east.

\begin{figure}
	\begin{center}
	\includegraphics[width=8cm]{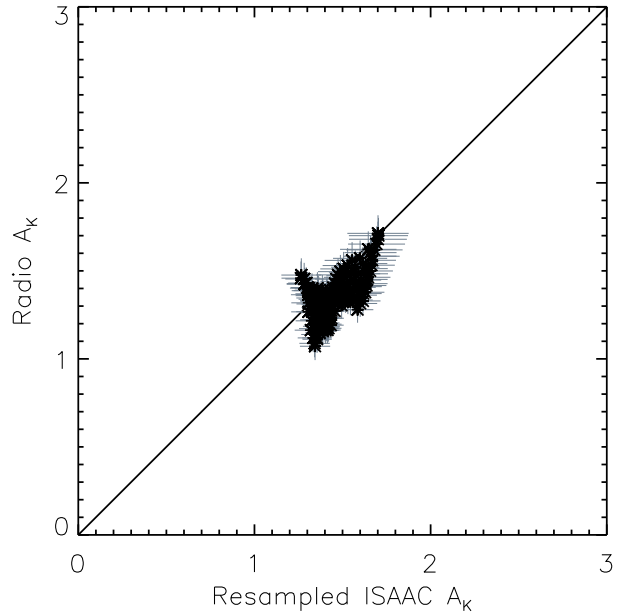}
	\end{center}
	\caption{ISAAC $A_K$ (resampled to the same resolution as the radio data) against Radio $A_K$. The cluster of points are those from the radio B H~{\sc ii} region.   }
	\label{fig9}
\end{figure}

In Figure~\ref{fig9}, the correlation between the extinction derived with the NIR method against that derived with the radio data is shown (after discarding points with Pa$\beta$ S/N $<$ 3). The distribution of points is consistent with a 1:1 correlation at low extinction (radio B), with both ISAAC and Radio methods yielding $A_K$ $\sim$ 1.5. The valid NIR method points do not cover the high extinction radio A source as there is very little Pa$\beta$ flux in this region.

\begin{figure}
	\begin{center}
	\includegraphics[width=8cm]{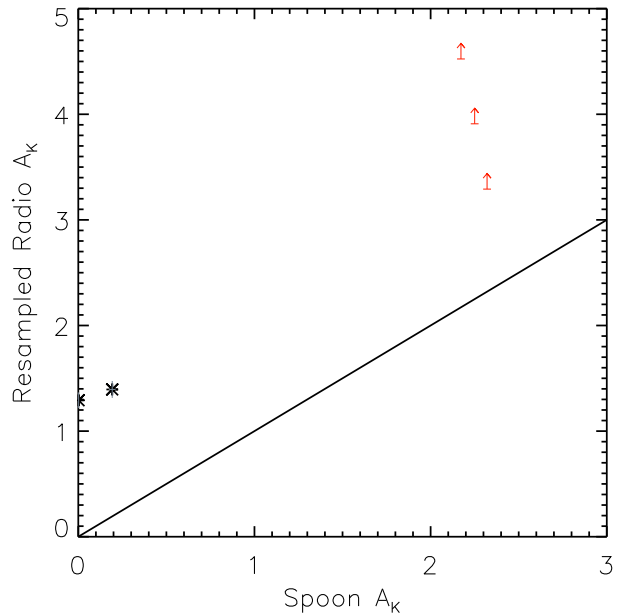}
	\end{center}
	\caption{Spoon $A_K$ against Radio $A_K$ (resampled to Spitzer/IRS SL pixel size: 1.8\arcsec). High extinction points are associated with the pixels associated with radio A, whilst the low extinction points are from radio B. }
	\label{fig11}
\end{figure}
Figure~\ref{fig11} shows the correlation between the extinction derived from the radio data against that derived using the Spoon method. The high extinction points near radio A have been included as lower limits (see later discussion). At low $A_K$ (radio B), the radio method presents a value of around 1.4, while the Spoon method finds a value very close to zero ($<$ 0.2).

\begin{figure}
	\begin{center}
	\includegraphics[width=8cm]{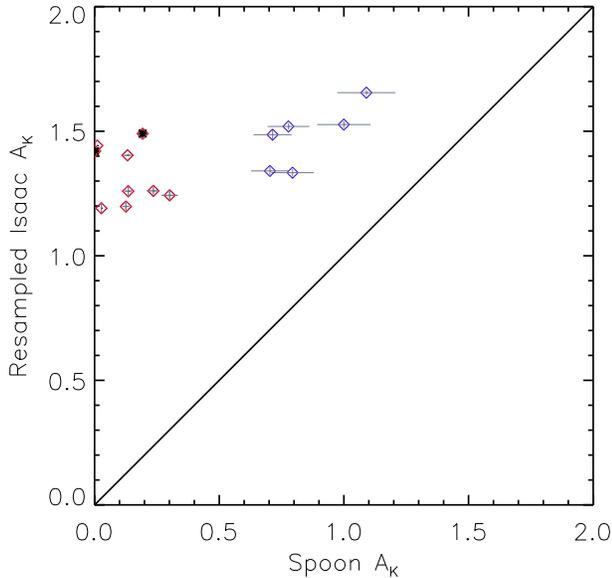}
	\end{center}
	\caption{Spoon $A_K$ against ISAAC $A_K$. The radio B pixels from Figure ~\ref{fig11} are marked as black crosses, the remaining points are included as blue or red diamonds corresponding to regions away from or associated with the radio B H~{\sc ii} region (i.e. within the southern low extinction region in Figure~\ref{fig_sil}) respectively.   }
	\label{fig10}
\end{figure}

In Figure~\ref{fig10}, the correlation between the extinction derived from the NIR method versus that derived using the Spoon method is shown. Figure~\ref{fig10} includes all of the data for the region surrounding IRAS 12063-6259 in Figure~\ref{fig10} as both the Spitzer/IRS map and the ISAAC data are valid away from the brightest regions of the source. In Figure~\ref{fig10}, the points associated with the radio B H~{\sc ii} region have been indicated (red points) and all possess lower Spoon extinction than those away from the radio B H~{\sc ii} region (blue points) as is already obvious from a cursory inspection of the spectra (c.f. Figure~\ref{fig6}). In Figure~\ref{fig10}, the S/N threshold for including points was increased to 5 to better display the discrepancy between the area of low silicate absorption near radio B and the higher absorption in the surroundings.
In principle the radio method is the most accurate at measuring the total extinction between the emitting source and the observer. The radio observations detect the total emission from the H~{\sc ii} region and so reflect the true shape of the ionized region, while at Br$\gamma$ the morphology is more complex and does not totally match that in the radio. Around radio B, the Br$\gamma$ and radio morphologies match and the extinction measured is the same as that of the NIR method. However near radio A, it is debatable whether any of the detected Br$\gamma$ flux emanates from the radio A H~{\sc ii} region at all. The morphology of the emission is certainly very different, indicating the possibility that this may be diffuse emission associated with IR 12063 or foreground emission. In that case, the extinction derived from the Br$\gamma$-radio continuum comparison provides a lower limit to the extinction of source A (and an upper limit to the extinction associated with the surrounding diffuse H~{\sc ii} component).
Near radio B, a consistent extinction is found with the NIR and the radio methods, but a discrepant extinction, of around A$_K$ $\sim$ 0.2, is found using the Spoon method. The emission morphology around radio B is very similar in the radio and NIR, so we can safely assume that the extinctions measured are the extinction between the radio B H~{\sc ii} region and the observer for the ISAAC and radio methods. The discrepant measurement arising from the Spoon method around radio B was checked using an independent measure of the silicate absorption feature: PAHFIT \citep{2007ApJ...656..770S}. 

\begin{figure}
	\begin{center}
	\includegraphics[width=8cm]{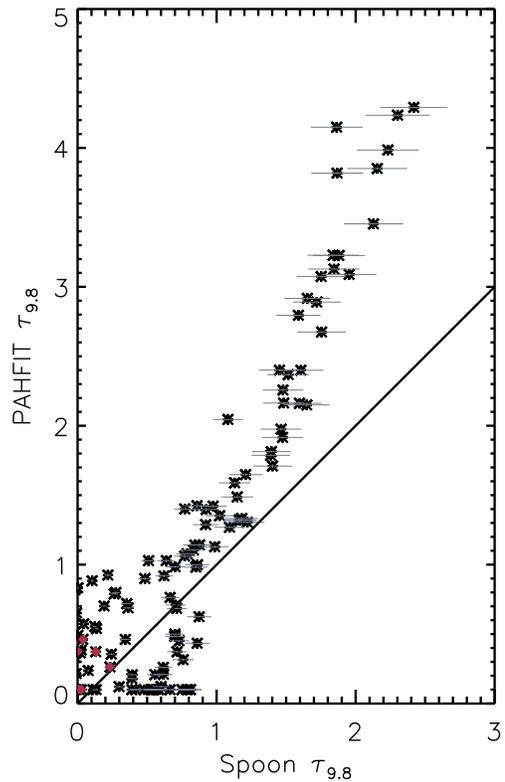}
	\end{center}
	\caption{Spoon $\tau_{9.8}$ against PAHFIT $\tau_{9.8}$. The Pixels from radio B have been indicated with filled red diamonds. The black line indicates a 1:1 correlation.  }
	\label{fig10a}
\end{figure}

Both methods agree that the pixels coincident with radio B, marked with red diamonds in Figure~\ref{fig10a}, have low silicate optical depths ($\tau_{9.8} < 0.5$) in agreement with the Spoon method, while the bulk of the rest of the points display much higher silicate absorption. At low optical depths, the points in Figure~\ref{fig10a} seem to agree, while at higher optical depths ($>1$) PAHFIT finds a higher optical depth (by around a factor of two) than the Spoon method. \citet{2008ApJ...676..304B} found the opposite trend in that they quote a higher range of values for the Spoon method rather than PAHFIT. 

PAHFIT was used to measure each pixel of the IRS-SL cube of IRAS 12063-6259, and the resulting correlation between the $\tau_{9.8}$ of PAHFIT against that of the Spoon method is shown in Figure~\ref{fig10a}. It should be noted that PAHFIT is intended to fit full IRS spectra (5--40 \micron), and large uncertainties result from using PAHFIT with only the IRS/SL range (5--14 \micron) because the longer wavelength continuum greatly aids the fitting process. However there exist no longer wavelength observations of IRAS 12063-6259 so the silicate absorption strength as measured by PAHFIT result from its application to only the IRS/SL range. In order to achieve a sensible extinction result from PAHFIT, it was necessary to impose a floor of $\tau_{9.8} = 0.01$ because it was found that PAHFIT frequently found fits to many of the pixels in the map with zero silicate absorption, even for pixels where the Spoon method found considerable optical depth (up to $\tau_{9.8}(\textrm{Spoon}) \sim 2$). Some of  Even after imposing this floor, there is still evidence of this effect in Figure~\ref{fig10a}, where there is a line of points at $\tau_{9.8}(\textrm{PAHFIT}) = 0.01$ and a range of Spoon values from zero to around 0.8. This implies that PAHFIT found an acceptable fit with effectively zero silicate absorption in spectra where the Spoon method detects optical depths of 0.8. From Figure~\ref{fig10a} we can also see the opposite effect, there are some pixels in which PAHFIT detects optical depths of up to 0.8 where the Spoon method detects zero optical depth, so it seems that under low optical depth conditions, both methods carry large uncertainties.

\begin{figure}
	\begin{center}
	\includegraphics[width=8cm]{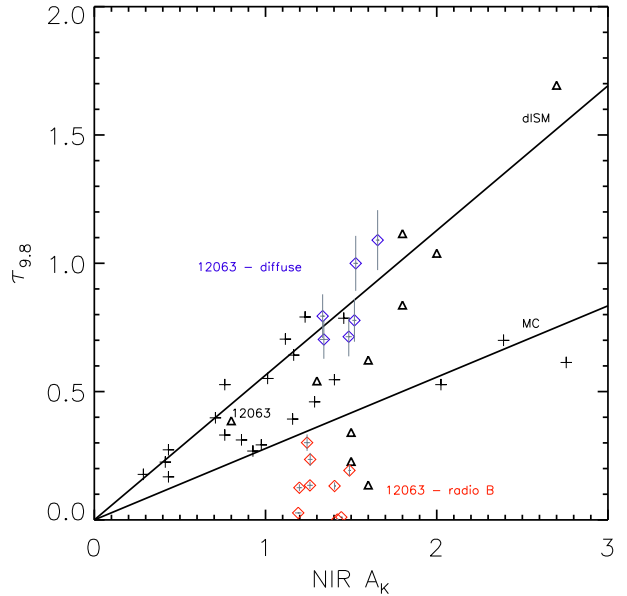}
	\end{center}
	\caption{Comparison of $\tau_{9.8}$ (Spoon) and NIR extinction with reference points representing the diffuse ISM and molecular clouds (black crosses, \citealt{2011ApJ...731....9C}), a sample of H~{\sc ii} regions observed with ISO-SWS (\citealt{2002A&A...381..571P}; black triangles, see text) and the measurements of the area around radio B and the diffuse emission from around IRAS 12063-6259 (red diamonds and blue diamonds respectively, the same points as in Figure~\ref{fig10}).  Best fit lines for ISM and molecular cloud sightlines are shown in black. }
	\label{figtau}
\end{figure}

In the diffuse ISM, the 9.8 \micron\ optical depth (as provided by the Spoon method) and extinction derived in the NIR have been found to follow a tight correlation (e.g. \citealt{2003dge..conf.....W} and references therein). Hints that there may be a different relationship between these quantities in some circumstances, i.e. possibly a shallower correlation, were found by \citet{1988MNRAS.233..321W}. Later studies confirmed this difference for molecular cloud sightlines and found that both the 9.8 \micron\ silicate optical depth, and the total optical depth at 9.8 \micron\ (continuum + silicates) correlate well with the NIR extinction, albeit with a shallower dependence for molecular clouds than the diffuse ISM (e.g. \citealt{2007ApJ...666L..73C}; \citealt{2009ApJ...693L..81M}; \citealt{2011ApJ...731....9C}; \citealt{2011A&A...526A.152V}). The data and correlations from these studies, along with that of IRAS 12063-6259 are shown in Figure~\ref{figtau}. None of the aforementioned studies found objects as extreme as radio B (NIR A$_K$ = 1.5, $\tau_{9.8}$ = 0.2). The areas of diffuse emission around the radio B H~{\sc ii} region are slightly above the general ISM trend (NIR A$_K$ $\sim$ 1.5, $\tau_{9.8}$ $\sim$ 1.0; blue points in Figure~\ref{fig10}). 

In Figure~\ref{figtau}, a selection of H~{\sc ii} regions have been included which were present in the \citet{2002AA...381..606M} sample and so have NIR A$_K$ measurements. The corresponding measurements of their silicate optical depth were found by applying the Spoon method\footnotemark to their ISO-SWS spectra \citep{2002A&A...381..571P}. The large ISO-SWS beam means that both the NIR A$_K$'s and silicate optical depths are spatial averages across each H~{\sc ii} region. In general the ISO-SWS H~{\sc ii} region sample appear along the ISM trend in Figure~\ref{figtau}. However three of them: IRAS 02219+6152, IRAS 10589-6034 and IRAS 18434-0242 (better known as G029.96-00.02), appear in roughly the same place as radio B in Figure~\ref{figtau}. IRAS 10589-6034 is part of a larger complex of very bright H$\alpha$ emission (saturated in the SHS) and also appears with similar morphology in each of the 2MASS bands. These points are much lower than either the molecular cloud or ISM trends. The three sources mentioned all have higher values of $\tau_{9.8}$ shown in the literature, (e.g. \citet{1978ApJ...221..797H} for IRAS 02219+6152) however, there is very little trace of absorption in the ISO-SWS spectra.

\footnotetext{In this case the Spoon method was not applied iteratively, in order to match the measurement process of other data shown in Figure~\ref{figtau}.}

From inspection of Figure~\ref{figtau}, it appears that H~{\sc ii} regions follow the ISM trend when spatially integrated but that under certain circumstances they can violate this trend and have low silicate absorption. In the case of IRAS 12063-6259, the radio B points appear to follow the molecular cloud trend while the diffuse surroundings follow the diffuse ISM trend. The three sources which appear near the radio B points in Figure~\ref{figtau} presumably represent objects with similar conditions to radio B, but without the surrounding deep silicate absorption that is seen in IRAS 12063-6259 as this would appear in the ISO-SWS spectrum. We note that the general picture is the same if PAHFIT silicate absorption measurements are used in place of the Spoon measurements, except that the points representing the diffuse surroundings of radio B fall well above the DISM trend. Included in Figure~\ref{figtau} is the ISO-SWS measurement of IRAS 12063-6259 (as indicated in the Figure), which was treated in the same way as the other ISO-SWS spectra. The ISO-SWS aperture used for this observation (shown in Figure~\ref{fig1}), includes both the bright southern part of IRAS 12063-6259 as well as the surrounding areas of high silicate extinction. The ISO-SWS observation of IRAS 12063-6259 (and by inference possibly the observations of the other H~{\sc ii} regions) represents a combination of both effects and as such the IRAS 12063-6259 point appears near, but below, the DISM trend in Figure~\ref{figtau}.

The physical interpretation of the differing relationship between $\tau_{9.7}$ and the NIR extinction is usually attributed to the effects of grain growth in molecular clouds. Grain coagulation causes the NIR opacity to rise relative to the 9.8 \micron\ silicate absorption because silicates are more easily coagulated than carbonaceous material. While \citet{2011A&A...526A.152V} found that it was very difficult to reproduce the observed relationship between $\tau_{9.7}$ and the NIR extinction without significantly altering the shape of the 9.8 \micron\ feature profile, the shape of the 9.8 \micron\ silicate absorption feature in the vicinity of radio B cannot be directly measured as there is so little absorption (see Figure~\ref{fig6}). The other major possibility which could solve the discrepancy between the NIR extinction and the silicate optical depth is that properties of the NIR extinction, generated by carbonaceous materials is different in the vicinity of radio B. This possibility can be ruled out on the grounds that the radio and NIR extinction measurements appear to be consistent with standard NIR extinction laws (e.g. Eq.~\ref{eq:AK}). 

The preceding discussion dramatically alters the picture of how the extinction maps can be applied to dereddening the mid-IR cube. It appears that each extinction map is valid in somewhat different regions. The NIR extinction map, for example is only valid around radio B, where there is sufficient S/N to make an accurate determination of the extinction. The radio extinction map is also only accurate around radio B, and over a much smaller area than the NIR map (because the radio observations are not sensitive to structure on larger scales). The Spoon and PAHFIT maps appear to not be trustworthy in the vicinity of radio A or B, as the previous discussion has shown that the ratio of silicate absorption to NIR extinction around both is much lower than would be expected.

\subsection{PAH Band Ratio Variations and Extinction}

In order to investigate the effects of extinction on the ratios of the major PAH bands, the IRS cube (as described in Section~\ref{sec:IRS}) was dereddened using the \citet{2006ApJ...637..774C} prescription for the extinction in the mid-IR relative to $A_K$ using the extinction maps described in the previous section. The regions for which the extinction maps are not thought to be trustworthy have been removed and are not presented in the following analysis. 

The continuum subtraction and flux measuring process as described in Section~\ref{sec:IRS} was repeated for cubes dereddened using the Spoon and NIR methods in the areas of their validity. In this section the unaltered observations will be referred to as the `observed' data, while the dereddened data will be referred to by the extinction map which was used to create it, e.g. Spoon or NIR.

Following the discussion in Section~\ref{sec:extdisc}, the extinction maps were used for dereddening in areas where they are valid. Therefore, in the case of the Spoon extinction map, the area around both radio sources is likely untrustworthy and has been masked out. For the NIR map, only the regions where the Pa$\beta$ signal to noise is above three at native resolution are included. The radio extinction map has not been used for dereddening purposes as it covers a very small area and would produce very few dereddened points.

\begin{figure}
	\begin{center}
	\includegraphics[width=8cm]{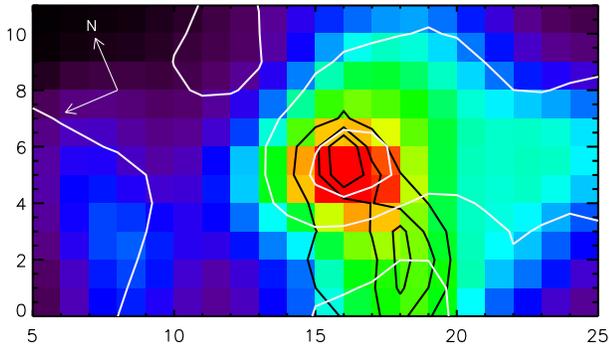}
	\end{center}
	\caption{Map of the 6.2 \micron\ PAH emission. Black contours represent 4.8 GHz brightness temperature 50, 100, 150 K; white contours represent the silicate absorption as measured by the Spoon method, peaking at radio A. The $x$ and $y$ axes refer to positions in pixels on the IRS cube (1.8\arcsec\ per pixel). }
	\label{fig12}
\end{figure}
 
\begin{figure}
	\begin{center}
	\includegraphics[width=8cm]{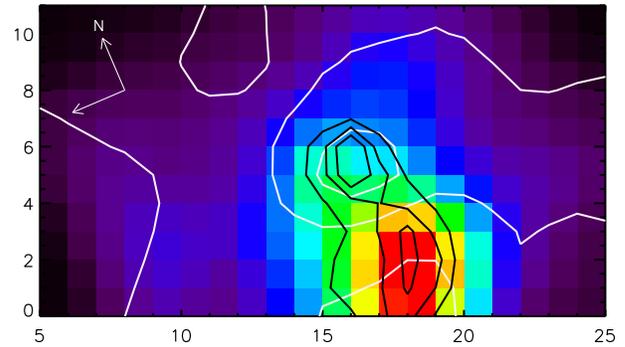}
	\end{center} 
	\caption{Map of the 8.6 \micron\ PAH emission. Black contours: 4.8 GHz brightness temperature 50, 100, 150 K; white contours represent silicate absorption as measured by the spoon method, peaking at radio A. The $x$ and $y$ axes refer to positions in pixels on the IRS cube (1.8\arcsec\ per pixel).  }
	\label{fig12a}
\end{figure}

The spatial distributions of the PAH bands (shown in Figures~\ref{fig12}~and~\ref{fig12a}) show strong spatial variations which seem to correlate with extinction. The 6.2 and 7.7 \micron\ bands peak in absolute strength near Radio A as well as being very similar morphologically (see Figure~\ref{fig12} for 6.2 \micron\ map along with radio and silicate absorption contours). In both maps the radio A peaks are all slightly offset to the south of radio A, possibly attributable to the large gradient in extinction in this region. They also show additional emission extending west towards the edge of the map. The 6.2 and 7.7 \micron\ maps also display isolated weak emission east of Radio A. The 8.6 and 11.2 \micron\ bands peak around Radio B and also share the same morphology, with some extended emission towards Radio A. Figure~\ref{fig12a} shows the morphology of the 8.6 \micron\ emission, as compared to the radio emission and the strength of silicate absorption. A comparison of Figures~\ref{fig12} and ~\ref{fig12a} shows the effects of the silicate absorption band on PAHs in that the morphologies are dramatically different. The area around radio A, associated with the highest extinction values, has strong 6.2 and 7.7 \micron\ and weak 8.6 and 11.2 \micron\ emission. 

In general, ratios of PAH band intensities (i.e. I$_{6.2}$/I$_{11.2}$) are taken to remove any influence that distance, abundance or variations in the intrinsic strength of the whole PAH spectrum may play on the relationships between the bands. This allows unbiased inspection of the relative strengths of different bands. 

\begin{figure}
	\begin{center}
	\includegraphics[width=8cm]{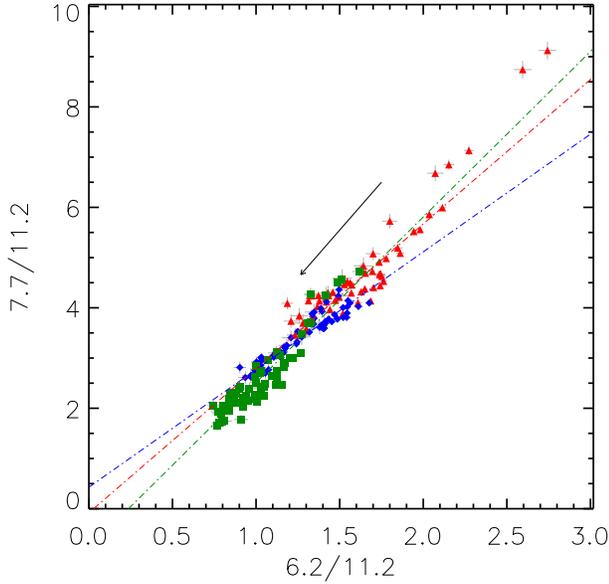}
	\end{center}
	\caption{  I$_{6.2}$/I$_{11.2}$ plotted against I$_{7.7}$/I$_{11.2}$ for the pixels surrounding Radio A and B. The red triangles represent the observed ratios while the blue diamonds represent the same data after dereddening using the Spoon extinction map and the green squares are similar using the ISAAC extinction map. Best fit gradients for each set of points are shown in the same color. A dereddening vector corresponding to $A_K$ = 1 is shown in black. }
	\label{fig14}
\end{figure}

\begin{figure}
	\begin{center}
	\includegraphics[width=8cm]{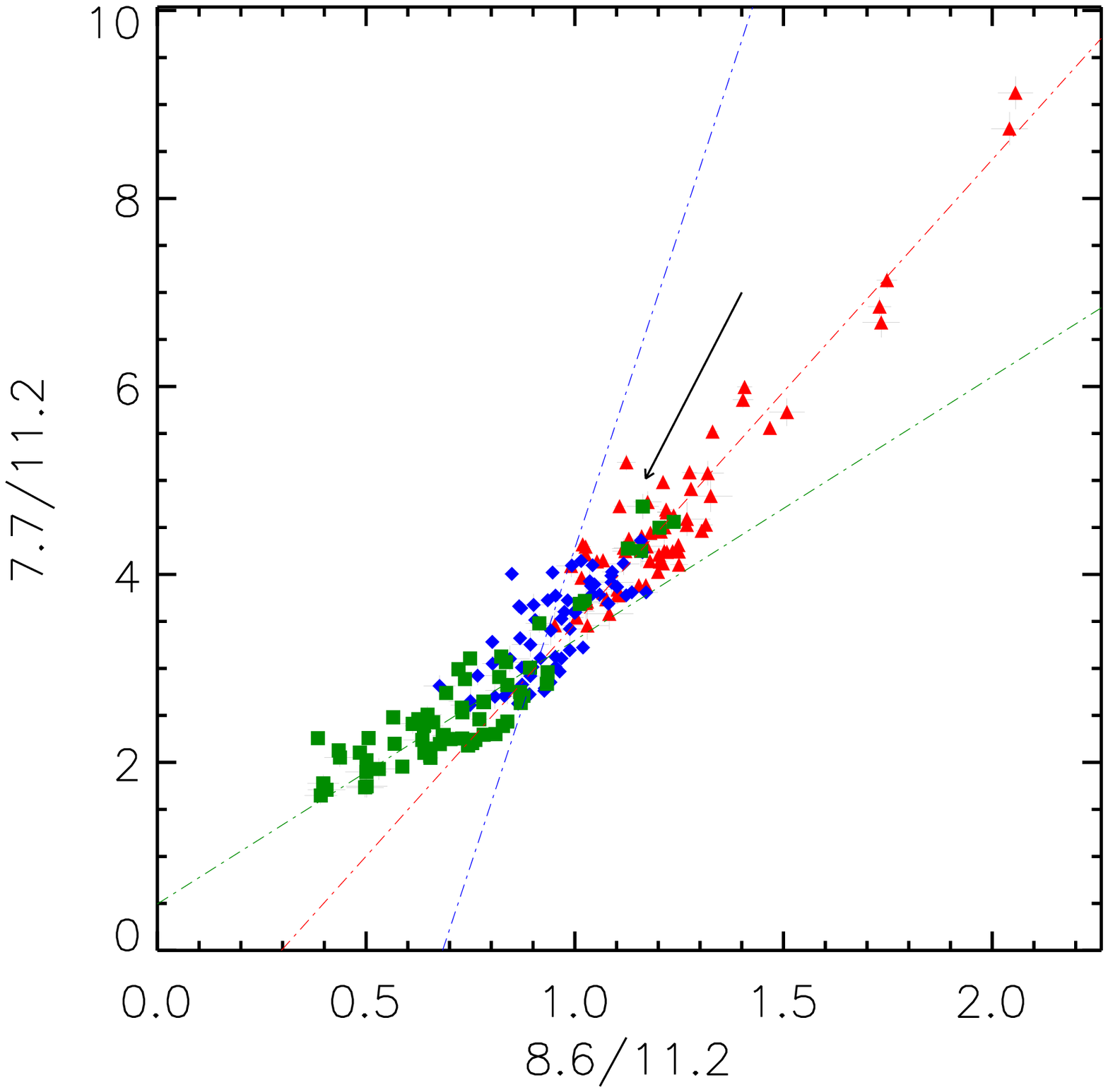}
	\end{center}
	\caption{  I$_{8,6}$/I$_{11.2}$ plotted against I$_{7.7}$/I$_{11.2}$ for the pixels surrounding Radio A and B. The red, blue and green points are as in Figure~\ref{fig14}. Best fit gradients for each set of points are shown in the same color. A dereddening vector corresponding to $A_K$ = 1 is shown in black.    }
	\label{fig16}
\end{figure}

\begin{figure}
	\begin{center}
	\includegraphics[width=8cm]{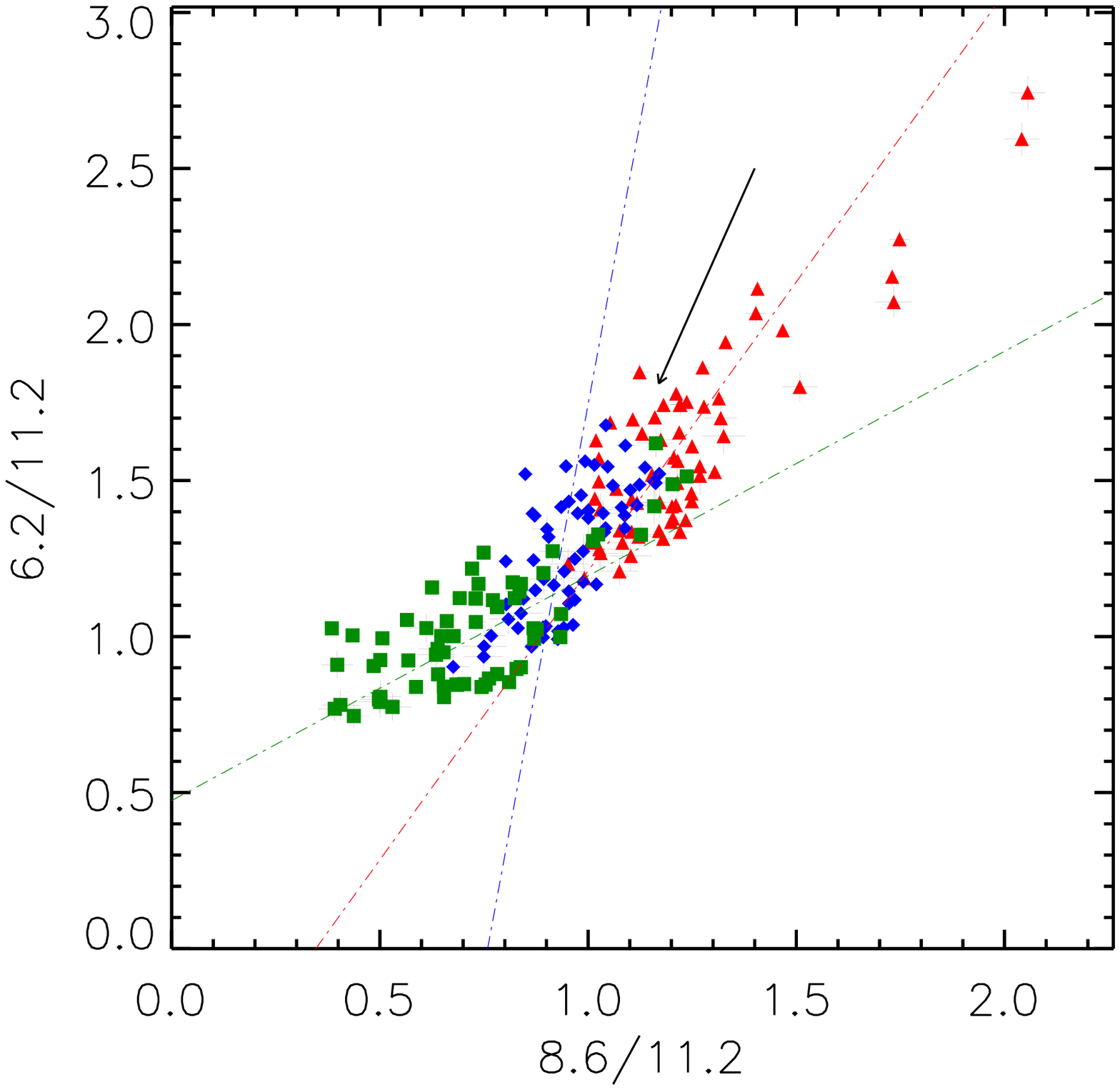}
	\end{center}
	\caption{  I$_{8.6}$/I$_{11.2}$ plotted against I$_{6.2}$/I$_{11.2}$ for the pixels surrounding Radio A and B. The red, blue and green points are as in Figure~\ref{fig14}. Best fit gradients for each set of points are shown in the same color. A dereddening vector corresponding to $A_K$ = 1 is shown in black.   }
	\label{fig17}
\end{figure}

\begin{table}
\caption{\label{table_PAHparms}Parameters of PAH Intensity Ratio Correlations}
\begin{center}
\begin{tabular}{l l l p{1.4cm} p{1.4cm} }
Data & Slope & Intercept & Slope Y(0)=0 & Correlation Coefficient  \\
\hline
\\
\multicolumn{5}{l}{$I_{6.2}$/$I_{11.2}$ vs $I_{7.7}$/$I_{11.2}$}\\
Observed     &  3.19$\pm$0.06 & -0.60$\pm$0.11 & 2.84$\pm$0.01  & 0.961\\
Spoon        &  2.27$\pm$0.06 &  0.53$\pm$0.07 & 2.72$\pm$0.01  & 0.947\\
NIR          &  3.66$\pm$0.26 & -1.19$\pm$0.25 & 2.47$\pm$0.03  & 0.914\\
NGC 2023$^a$ &  2.04$\pm$0.02 & -0.29$\pm$0.03 & 1.89$\pm$0.01  & 0.967  \\
M17$^b$      &                &                & 1.96           & 0.81 \\
\\
\multicolumn{5}{l}{$I_{8.6}$/$I_{11.2}$ vs $I_{7.7}$/$I_{11.2}$}\\
Observed     &   5.24$\pm$0.15 & -1.71$\pm$0.19 & 3.90$\pm$0.03  & 0.954\\
Spoon        &  17.01$\pm$3.37 &-12.77$\pm$3.17 & 3.42$\pm$0.03  & 0.636\\
NIR          &   2.17$\pm$0.18 &  0.97$\pm$0.13 & 3.57$\pm$0.08  & 0.871\\
NGC 2023$^a$ &   4.40$\pm$0.06 &  1.17$\pm$0.03 & 6.31$\pm$0.04  & 0.953 \\
M17$^b$      &  -              &                & 11.91          & 0.33 \\
\\
\multicolumn{5}{l}{$I_{8.6}$/$I_{11.2}$ vs $I_{6.2}$/$I_{11.2}$}\\
Observed    &   1.66$\pm$0.07 & -0.38$\pm$0.09 & 1.37$\pm$0.01  & 0.880\\
Spoon       &  10.19$\pm$2.82 & -8.40$\pm$2.65 & 1.25$\pm$0.01  & 0.593\\
NIR         &   0.41$\pm$0.08 &  0.70$\pm$0.06 & 1.43$\pm$0.05  & 0.676\\
NGC 2023$^a$&   2.13$\pm$0.03 &  0.72$\pm$0.03 & 3.44$\pm$0.02  & 0.949 \\
\end{tabular}
\bigskip
\end{center}
$^a$: Peeters et al., 2013 in prep.\\
$^b$: \citet{2008ApJ...679..310G}
\end{table}

The most commonly quoted correlation, I$_{6.2}$/I$_{11.2}$ vs I$_{7.7}$/I$_{11.2}$, is presented in Figure~\ref{fig14}).  Similar trends are seen for both I$_{8.6}$/I$_{11.2}$ vs I$_{7.7}$/I$_{11.2}$ (Figure~\ref{fig16}) and I$_{8.6}$/I$_{11.2}$ vs I$_{6.2}$/I$_{11.2}$ (Figure~\ref{fig17}) in the observed data. Fits to each of these correlations (along with the dereddened correlations which are subsequently presented) are presented in Table~\ref{table_PAHparms}. Separate gradients for each correlation are provided which pass through the origin or are allowed to vary their y-intercepts.

In each of Figures~\ref{fig14},~\ref{fig16}~and~\ref{fig17} dereddened points have also been included. For the I$_{6.2}$/I$_{11.2}$ vs I$_{7.7}$/I$_{11.2}$ correlation, the points with the highest ordinate values correspond to the points of highest extinction around radio A. This collection of points (I$_{7.7}$/I$_{11.2}$ $>$ 5) clearly displays a different slope than the rest of the points for the low extinction parts of the source. The dereddened points in Figure~\ref{fig14} remove the `break' in the slopes evident in the observed data. 

The general effect of extinction of the I$_{6.2}$/I$_{11.2}$ vs I$_{7.7}$/I$_{11.2}$ correlation is most straightforward to explain: the 11.2 \micron\ PAH band is affected by extinction from the 9.8 \micron\ silicate feature to a greater degree than the 6.2 or 7.7 \micron\ bands. Therefore the effect of extinction is to preferentially reduce the 11.2 \micron\ PAH flux and increase both the I$_{6.2}$/I$_{11.2}$ and I$_{7.7}$/I$_{11.2}$ ratios. The major effect of dereddening this correlation is to decrease each of the ratios and as such move all points towards the origin. 

Comparing the data in Table~\ref{table_PAHparms} for IRAS 12063-6259 to that of NGC 2023, which was measured in the same way with the same software (Peeters et al., 2013, in prep.), it is clear that there is no agreement between any of the parameters of the correlations between the PAH bands as observed for IRAS 12063-6259 and NGC 2023.  Included in Table~\ref{table_PAHparms} are the correlations derived by \citet{2008ApJ...679..310G} for the M17 galactic star formation complex, which is the only galactic H~{\sc ii} region in the \citet{2008ApJ...679..310G} sample. The M17 correlation for I$_{6.2}$/I$_{11.2}$ vs I$_{7.7}$/I$_{11.2}$ gives a lower value than the 2.84\footnotemark\ found for the non-dereddened IRAS 12063-6259 data. The values found for IRAS 12063-6259 are lowered by the dereddening process to as low as 2.42, however this still significantly differs. Nevertheless, even after dereddening a tight correlation exists the the IRAS 12063-6259 data.

\footnotetext{\citet{2008ApJ...679..310G} quote only gradients for correlations which pass through the origin, so the quoted gradient is for the IRAS 12063-6259 correlation defined in the same way.}

The other correlations listed in Table~\ref{table_PAHparms} also disagree with those for NGC 2023 and M17, however in these cases the effect of dereddening, as discussed above, does not narrow the gap between the IRAS 12063-6259 data and that of the other objects. The two dereddening methods employed twist the best fit correlations in opposite ways, i.e. for the map dereddened using the Spoon method the gradient increases drastically (by a factor of 3--6). While the NIR dereddened data yields a shallower best fit, giving reductions in the gradients by a factor of 2--3. From inspection of the correlation plots (Figures~\ref{fig16} and \ref{fig17}) this appears to be because the Spoon method collapses the variation in the I$_{8.6}$/I$_{11.2}$ ratio to around one, while the NIR method gives a range from around 0.2 -- 1.2.

\begin{figure}
	\begin{center}
	\includegraphics[width=8cm]{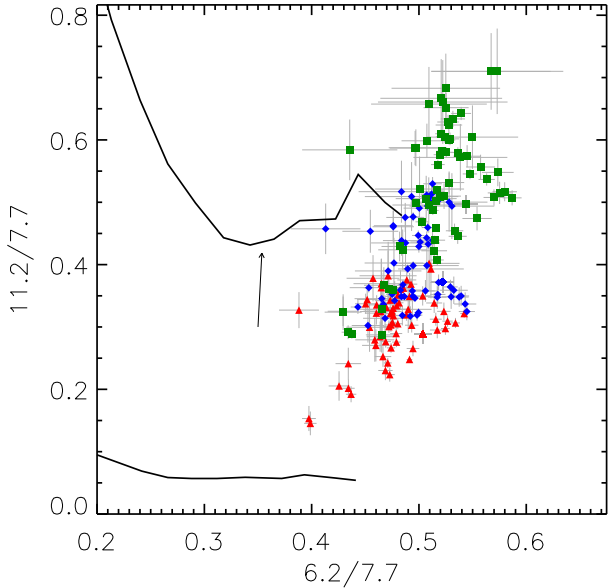}
	\end{center}
	\caption{  I$_{6.2}$/I$_{7.7}$ vs I$_{11.2}$/I$_{7.7}$ for the area around IRAS 12063-6259. The red, blue and green points are as in Figure~\ref{fig14}. The two lines represent ionized (lower) and neutral (upper) PAH tracks for a variety of PAH molecule sizes (increasing towards the left) \citep{2001ApJ...551..807D}. A dereddening vector corresponding to $A_K$ = 1 is shown in black.   }
	\label{fig13}
\end{figure}

In Figure~\ref{fig13}, another commonly examined correlation I$_{6.2}$/I$_{7.7}$ vs I$_{11.2}$/I$_{7.7}$ is shown, which is considered to be a useful diagnostic of PAH size and ionization \citep{2001ApJ...551..807D}. The tracks for ionized and neutral PAHs at varying $N_c$ are adopted from \citet{2001ApJ...551..807D}. The observed IRAS 12063-6259 points fall between the tracks, indicating a mixture of ionized and neutral PAHs, albeit at the lowest end of the $N_c$ values considered ($N_c$ = 16!). However it is clear that the dereddening process has significantly altered the I$_{11.2}$/I$_{7.7}$ ratio, as might be expected given that $^{A_\lambda}/_{A_K}$ at 11.2 \micron\ is higher than the other PAH bands. In fact, dereddening has increased the range of values of I$_{11.2}$/I$_{7.7}$ from $\sim$0.15 to around 0.3. The range of values in I$_{6.2}$/I$_{7.7}$ does not change significantly as the difference between $^{A_\lambda}/_{A_K}$ at 6.2 and 7.7 \micron\ are much smaller. 

\begin{figure}
	\begin{center}
	\includegraphics[width=8cm]{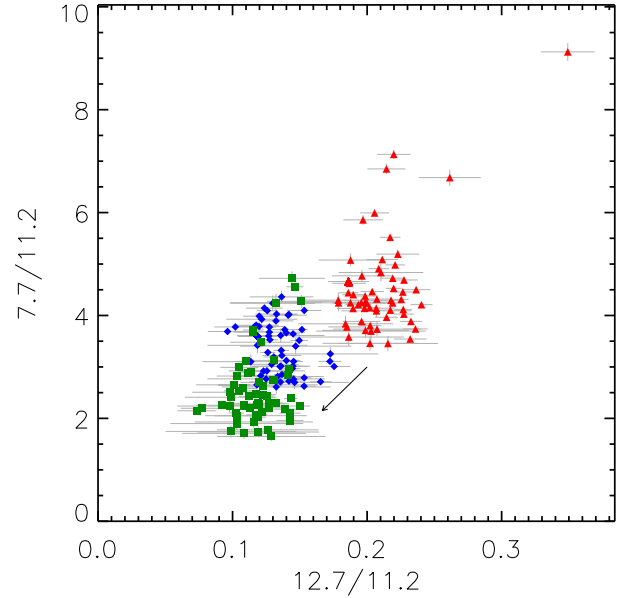}
	\end{center}
	\caption{  I$_{7.7}$/I$_{11.2}$ plotted against I$_{12.7}$/I$_{11.2}$ for the area around IRAS 12063-6259. The red, blue and green points are as in Figure~\ref{fig14}. A dereddening vector corresponding to $A_K$ = 1 is shown in black.  }
	\label{fig17a}
\end{figure}

In addition to the main bands, the 12.7 \micron\ band was measured as it should also be weakly affected by extinction. In Figure~\ref{fig17a} the observed I$_{6.2}$/I$_{11.2}$ vs I$_{12.7}$/I$_{11.2}$ correlation is shown, along with the dereddened versions as in previous figures. The ratio of I$_{12.7}$/I$_{11.2}$ is of particular note as its variation decreases the most after dereddening. Dereddening via the Spoon method in particular reduces variation in the I$_{12.7}$/I$_{11.2}$ ratio to almost zero.

\section{Discussion}

\subsection{The Morphology and Geometry of IRAS 12063-6259}

The somewhat conflicting extinction maps suggest a geometry for IRAS 12063-6259: radio A is actually a deeply embedded, highly extinguished H~{\sc ii} region separated from the H~{\sc ii} region around radio B. The PAH bands are seen to sharply peak at the same position as radio A, which strongly argues for most of the emitting PAHs being located in the PDR around the radio A H~{\sc ii} region and not associated with the extended PAH emission evident in the Figure~\ref{fig12}. In this case, the radio A PAH emission should suffer from a similar level of extinction as the ionized part of the H~{\sc ii} region, yet the MIR spectra reveal only modest silicate absorption. The fact that the Spoon method finds a much lower extinction (at least a factor of two) than the lower limit given by the radio method is possibly another instance of the expected relationship between the NIR extinction and the silicate absorption breaking down.

The radio B source, on the other hand, suffers from a similar problem which we must tentatively attribute to silicate coagulation. The non zero NIR extinction and near zero silicate optical depth are difficult to explain using other mechanisms especially as the NIR extinction around radio B is consistent with standard extinction laws. Radio B appears to have cast off its extinguishing material (possibly due to it being a small cluster of stars as suggested by \citealt{2003AA...407..957M}). It is also possible that radio B is simply older and has cast off its natal cloud. Given the lack of extinguishing material on our line of sight, it seems that the stars making up radio B are the likely cause of the extended PAH emission surrounding IRAS 12063-6259.

\subsection{Methods of Measuring Extinction}

Here we discuss the possible ramifications of our findings regarding the various methods of measuring extinction. It is important to note that each method, in isolation, presents a seemingly reliable measurement of extinction. It is only when compared to other independent measures of extinction that inconsistencies emerge.


The Spoon method is commonly used to classify external galaxies (e.g. \citealt{2007ApJ...654L..49S}). In this role it provides a self-consistent way of determining the degree of silicate absorption, albeit not a very accurate in terms of the total extinction as it seems to systematically underestimate the degree of silicate optical depth (as shown in Section~\ref{sec:sil_abs} where we developed a method of compensating for this effect). In addition, it does not seem to be linked to the overall extinction, as measured by the radio method, as has also been concluded from completely independent methods to measure the extinction associated with molecular cloud materials (e.g. \citealt{2007ApJ...666L..73C}; \citealt{2009ApJ...693L..81M}; \citealt{2011ApJ...731....9C}; \citealt{2011A&A...526A.152V}). 

Spatial extinction maps also show that different regions of a small object can have vastly different extinction properties. If we regard the ISO-SWS spectra of IRAS 12063-6259 as an average of the whole object, it shows some silicate absorption as well as NIR extinction. The Spitzer/IRS observations disentangle these two effects and show that they are spatially distinct on scales of the 0.4 pc separation between the two radio sources. In the case of IRAS 12063-6259, this small separation is resolved, for the usual application of the Spoon method (galaxy classification) variations on this scale will be entirely invisible.  

\subsection{PAH Variations and Extinction}

The primary argument made by \citet{2008ApJ...676..304B} was that because the variations in PAH band ratios were lower than the variations one would expect due to variations in the extinction; one could not reliably ascribe these variations to any other cause. \citet{2008ApJ...676..304B} made this assertion based on the scatter in their plot of I$_{6.2}$/I$_{7.7}$ against I$_{11.2}$/I$_{7.7}$ (which we show with our data in Figure~\ref{fig13}), which was rather close to the expected dereddening vector. For IRAS 12063-6259, it is observed that the range of points on this figure is maintained or even amplified upon dereddening (regardless of the extinction map chosen). This implies that the variations in the PAH band ratios are not due to variable extinction but are intrinsic. 

We find that correlations involving the 8.6 \micron\ band are highly affected by extinction, as one might expect. However the combination of the band being weak and the second most heavily affected by extinction produce very scattered correlation plots (e.g. Figures~\ref{fig16} and \ref{fig17}) with low correlation coefficients. This probably also arises from the uncertainty on the wavelength profile of the 9.8 \micron\ silicate absorption feature which has been found to vary significantly \citep{2011A&A...526A.152V}. The different extinction maps used to deredden the IRS cube have the largest effect on the 8.6/11.2 PAH ratio. The data dereddened using the Spoon map shows much lower dispersion than that dereddened using the NIR map. There are three potential explanations for this effect, it could be a reflection of the NIR extinction map being noisier than the Spoon map, although if this were the case it would be expected that the NIR points on the correlation plots would have a higher scatter without preferring any specific direction, which is not observed. Secondly, it could be that the method of using an extinction law to measure extinction and then dereddening using the same extinction law introduces biases which cause the Spoon dereddened points to cluster. Lastly, it could be that the Spoon method systematically fails at low $A_K$, in this case the points at low extinction do not move at all (as for low NIR $A_K$ the Spoon method seems to find $A_K$ = 0) while the high extinction points move towards them - decreasing the scatter. However, the agreement between the Spoon method and PAHFIT on the lack of absorption in radio B spectra seems to rule out the final possibility being a systematic error and suggest that instead, the 8.6/11.2 PAH ratio displays a lower variation than expected.

In general the PAH intensity correlations found for IRAS 12063-6259 are consistent with those of previous studies of H~{\sc ii} regions and also the reflection nebula NGC 2023. The range of values over which each PAH ratio was found to vary seems to be compressed for IRAS 12063-6259 as compared to other objects, with the exception of the 7.7 and 6.2 bands with respect to the 11.2. The 12.7 \micron\ PAH feature in particular displays very little variation as compared to other studies. From Figure~\ref{fig17a} it is clear that the variation in 12.7 relative to 11.2 is consistent with 12.7/11.2 being constant at a value of 0.2 (after dereddening the effect persists but the value is around 0.11 or 0.14 for cube dereddened using the NIR or Spoon extinction maps respectively). In contrast, the reflection nebula NGC 2023 shows a range of 0.2 -- 0.6 for 12.7/11.2 (Peeters et al. 2013; in prep.). In this sense the PAH population of IRAS 12063-6259 may not be the best test for the effect of extinction on PAHs as there is very little variation in some of the PAH intensity ratios. However there have been very few studies of the spatial distribution of PAHs and their properties around H~{\sc ii} regions.

\section{Summary \& Conclusions}

The spatial variations in extinction across IRAS 12063-6259 have been mapped using multiple methods. We then used these maps to deredden MIR observations and hence investigated the effects of the spatial extinction variations on PAH intensity variations. 

We obtained NIR, MIR and radio observations of the H~{\sc ii} region IRAS 12063-6259. The NIR observations were performed using the ISAAC instrument on the VLT and provided maps of the narrow band hydrogen recombination line emission. The MIR data were acquired using the Spitzer/IRS instrument in mapping mode and the radio observations were obtained using the ATCA array. These data allowed the derivation of several independent extinction maps. The NIR observations were used to make the first extinction map by comparing the ratio of the strengths of two hydrogen recombination lines to their intrinsic ratio as given by case B recombination theory. The second measure of extinction was calculated by using the continuum radio observations to find the intrinsic strength of the stronger of the two hydrogen lines measured in the NIR and then comparing the two. The final method, measuring the depth of the MIR silicate absorption feature, was measured in two ways, firstly using the Spoon method and subsequently using PAHFIT.

We showed that the Spoon method systematically underestimates the magnitude of the silicate absorption at 9.8 \micron\ in the absence of observations at wavelengths longer than 14.5 \micron. This effect originates from the secondary, shallower, silicate absorption feature which reduces the flux of the 14.5 \micron\ spline point used to derive the continuum by the Spoon method. We correct for this problem by implementing an iterative Spoon method which compensates for the lowered 14.5 \micron\ flux by dereddening the spectrum and then reapplying the Spoon method. The original Spoon method appears to underestimate the 9.8 \micron\ optical depth by around 40\% in high extinction regions without using this iterative technique.

The extinction measurements which use the ionized hydrogen observations (NIR, radio) tend to agree, while measurements based on the 9.8 \micron\ silicate absorption feature are in disagreement. Different methods of measuring the MIR silicate absorption yielded different results, with the Spoon method finding much lower absorption than PAHFIT (although the PAHFIT measurements were performed using only short wavelength IRS data). Comparing the silicate absorption to the NIR extinction showed that the dust composition varies on different sightlines to IRAS 12063-6259, being more like a molecular cloud on the radio B sightline and consistent with general ISM trends away from the two radio sources. Prompted by this discrepancy, ISO-SWS H~{\sc ii} region spectra \citep{2002AA...381..606M} were measured in the same way and this effect was also found for the H~{\sc ii} region IRAS 10589-6034, while other H~{\sc ii} regions were consistent with the ISM relationship between silicate absorption and NIR extinction. In addition, the radio extinction method showed a much higher than expected extinction compared to the silicate absorption for the radio A source. The silicate absorption observed in the MIR does peak around radio A, but only appears to account for half of the total extinction. We conclude that the spatial variations in extinction appear to include not only the magnitude of the extinction but also the dust composition along each sightline.

The two radio sources in IRAS 12063-6259 reported by \citet{2003AA...407..957M} possess remarkably different properties which have proven difficult to satisfactorily explain. Radio A, seems to be a heavily extinguished object with an A$_K$ greater than 4, yet it is ostensibly encompassed in a larger surrounding H~{\sc ii} region. There is a `bar' of extinction which appears to be related to IRAS 12063-6259 as there is diffuse emission associated with IRAS 12063-6259 emanating from this region. Radio B, which \citet{2003AA...407..957M} suggest could be multiple stars, is coincident with the brightest part of the H~{\sc ii} region and seems to have a much lower total extinction of around A$_K$ $\sim$ 1.5 (of which very little is generated by silicates).

The Spitzer/IRS spectral cube of IRAS 12063-6259 was dereddened using the extinction maps and the spectral features measured for each pixel in both the dereddened cubes as well as in the originally observed cubes. The standard correlation plots were then presented for both the observed and dereddened data. 

Correcting for extinction changes the PAH correlations, in most cases altering the gradients, moreover the range of each particular PAH ratio can also change substantially. Any correlation involving the 8.6 \micron\ band tends to be degraded by both noise and the relative uncertainty in the shape of the blue wing of the 9.8 \micron\ silicate absorption feature. In IRAS 12063-6259, correcting for extinction using the Spoon method removes variation in the 12.7 / 11.2 PAH band ratio. In general though, the dereddening process does not seem to totally remove variations in the PAH ratios as was suggested by \citet{2008ApJ...676..304B}.

\section*{Acknowledgments}

DJS thank to William Choi and Saied Sorkhu for their assistance in reducing and measuring the Spitzer/IRS observations. Much of the work in this paper relies on the inclusion of radio observations of IRAS 12063-6259 provided by N. Mart\'{i}n-Hern\'{a}ndez.

DJS and EP acknowledge support from an NSERC Discovery Grant and an NSERC Discovery Accelerator Grant. 

Studies of interstellar chemistry at Leiden Observatory are supported through advanced-
ERC grant 246976 from the European Research Council, through a grant by the Dutch
Science Agency, NWO, as part of the Dutch Astrochemistry Network, and through the
Spinoza premie from the Dutch Science Agency, NWO.

\bibliographystyle{apj}

\end{document}